\documentclass[12pt]{article}

\usepackage{times}

\newenvironment{sciabstract}{%
\begin{quote} \bf}
{\end{quote}}

\newcounter{lastnote}
\newenvironment{scilastnote}{%
\setcounter{lastnote}{\value{enumiv}}%
\addtocounter{lastnote}{+1}%
\begin{list}%
{\arabic{lastnote}.} {\setlength{\leftmargin}{.22in}}
{\setlength{\labelsep}{.5em}}} {\end{list}}

\usepackage{hyperref}
\usepackage{amsmath}
\usepackage{subfigure}
\usepackage{graphicx}
\usepackage{ragged2e}
\usepackage{ulem}

\title{Capacity Enhancement with Meta-Multiplexing}

\author
{C. Ji$^{\dag}$$^\ast$ and R. Liu$^{\dag}$\\
\\
\normalsize{\small\it Kuang-Chi Institute of Advanced Technology, Shenzhen, China}\\
\\
\normalsize{$^{\dag}$ These authors contributed equally to this
work.}\\
\normalsize{$^\ast$ To whom correspondence should be addressed.}
\normalsize{E-mails: chunlin.ji@kuang-chi.org.}
 }

\date{}

\begin{document}


\baselineskip24pt


\maketitle


\begin{sciabstract}
Multiplexing services as a key communication technique to
effectively combine multiple signals into one signal and transmit
over a shared medium. Multiplexing can increase the channel
capacity by requiring more resources on the transmission medium.
For instance, the space-division multiplexing accomplished through
the multiple-input multiple-output (MIMO) scheme achieves
significant capacity increase by the realized parallel channel,
but it requires expensive hardware resources. Here, we present a
novel multiplexing methodology, named meta-multiplexing, which
allows ordinary modulated signals overlap together to form a set
of ``artificial'' parallel channels; meanwhile, it only requires
similar resources as ordinary modulation schemes. We prove the
capacity law for the meta-multiplexing system and disclose that
under broad conditions, the capacity of a single channel increases
linearly with the signal to noise ratio (SNR), which breaks the
conventional logarithmic growth of the capacity over SNR. Numerous
simulation studies verify the capacity law and demonstrate the
high efficiency of meta-multiplexing. Through proof-of-concept
hardware experiments, we tested the proposed method in
communication practices and achieved a spectral efficiency of 81.7
bits/s/Hz over a single channel, which is significantly higher
than the efficiency of any existing communication system.
\end{sciabstract}

\newpage


In the celebrated 1948 paper \cite{Shannon_1948}, Shannon derived
the seminal formula for the capacity of the additive linear white
Gaussian noise (AWGN) channel: $C=W\log_2(1+\mathrm{SNR})$, where
the capacity $C$ is the tight upper bound of the rate at which
information can be reliably transmitted over the channel, the
capacity $C$ is determined by the bandwidth $W$ and the ratio of
signal power to noise power SNR. People have made tremendous
efforts to approach the ideal capacity limit
\cite{Costello_Forney_2006}. The capacity of a single channel is
hard to break, except for channels with nonlinearity
\cite{Mitra_Stark_2001,Sorokina_Turitsyn_2014}. People also appeal
to the multiplexing techniques
\cite{Proakis_2001,Tse_Viswanath_2005} to increase the capacity of
a communication system by allocating more resources on the
transmission medium: MIMO communication systems (see Fig. S1 in
\cite{SOM} ) establish a spatial parallel channel between the
transmitter and receiver using multiple antenna pairs. The
space-division multiplexing through the parallel channel
significantly increases the capacity of MIMO systems. For
instance, in an ideal situation, the capacity of a MIMO system is
$N$ times larger than a single channel's capacity, where $N$ is
the smaller of the number of transmitter or receiver antennas.
However, MIMO requires extra expensive radio frequency
infrastructures and the availability of multiple independent
channels, which may limit its applications. This paper introduces
a novel multiplexing framework that incorporates ordinary
modulated signals to construct an artificial parallel channel for
a single physical channel without requiring much extra time or
frequency resources. The proposed system requires more computation
efforts but leads to a dramatic increase in the capacity. Under
surprisingly broad conditions, such as various parameter setting
for communication systems or even with interference from
neighboring frequency bands, the system possesses a region in
which the capacity increases linearly with the SNR. We named the
system the meta-multiplexing, which is aligned to the
``artificial'' material ``metamaterial'' \cite{Liu_Ji_09}, as both
utilize ordinary elements to form an ``artificial'' entirety and
achieve extraordinary performance.

%
Recall that the enhanced capacity of the MIMO system is
accomplished by the realized parallel channel derived from the
channel matrix (see Fig. S1 in \cite{SOM}). However, the channel
matrix is leverage on the physical scattering propagation
environment (independent uncorrelated spatial paths) and is
difficult for humans to manipulate. In the meta-multiplexing, we
construct the parallel channel artificially: assuming that the
data is transmitted by symbols with symbol time $T$, Fig. 1A shows
the naive idea of the meta-multiplexing scheme, in which we
accelerate and parallel the transmission to $K$ streams and
introduce a carefully designed time delay $T/K$ for each chain of
the data stream. The multiple delays between different data
streams are predefined; thus, the signals can be superposed
together and propagate over a single physical channel, whereas at
the receiver, unlike MIMO, we can decode them without requiring
multiple antennas to capture different copies of the signal. We
can further simplify the structure and process the multiplexing in
the digital part as shown in Fig. 1B, which further reduces the
cost of the implementation. We illustrate the meta-multiplexing
process through a simple example shown in Fig. 2A. The binary
phase-shift keying (BPSK) signal is paralleled and transmitted
with a delay of $T/K$ for each stream. The summation of energy
levels at each time slot $T/K$ produces the multiplexed signal,
which is actually a waveform varying in one symbol time.
Straightforwardly, the multiplexing process is equivalent to a
convolution process in which the BPSK signal is convoluted with
the rectangle waveform. Moreover, the pulse shaping filter
$\textbf{h}(t)$ can be further extended to represent all of the
impulse responses of the entire communication system, including
the pulse shaping filter, the channel impulse response and the
matched filter if available. Intuitively, the benefit of this
scheme is that the data stream has been accelerated $K$ times and
transmitted in $K$ parallel chains, and as a consequence, the
information is carried by a novel format of waveform without
requiring a great amount of extra transmission resources: no
multiple antenna, no or limited extra bandwidth (discussed later)
and only a negligible extra transmission time $(K-1)/K*T$.

To explore the capacity benefits of the meta-multiplexing
precisely, we derive the capacity law for this communication
scheme. As shown in Fig. 1C, the system can be represented by the
matrix format $\mathbf{y}=\mathbf{H}\mathbf{x}+\mathbf{z}$, where
vector $\mathbf{x}$ and $\mathbf{y}$ denotes the transmitted and
received signal respectively, and $\mathbf{z}$ denotes the white
Gaussian noise with variance $N$; $\mathbf{H}$ is a matrix in
which each row is the impulse response coefficient $\mathbf{h}(t)$
shifted with a delay of $T/K$. By applying a singular value
decomposition, $\mathbf{H}=\mathbf{U} \mathbf{\Lambda}
\mathbf{V}$, with unitary matrix $\mathbf{U}$ and $\mathbf{V}$,
and diagonal matrix $\mathbf{\Lambda}$, the matrix expression of
the communication system is reformed as
$\tilde{\mathbf{y}}=\mathbf{\Lambda}
\tilde{\mathbf{x}}+\tilde{\mathbf{z}}$, where $\tilde{\mathbf{x}}$
, $\tilde{\mathbf{y}}$, and $\tilde{\mathbf{z}}$ are coordinate
transformations of $\mathbf{x}$, $\mathbf{y}$, and $\mathbf{z}$
\cite{Tse_Viswanath_2005,SOM}. The reformed expression clearly
shows that the transmission system is equivalent to a parallel
channel \cite{Tse_Viswanath_2005,Cover_Thomas_2006}. Thus, by
following the deduction of the capacity of the parallel channel
\cite{Cover_Thomas_2006}, the capacity of the meta-multiplexing
system can be obtained: $C=\frac{1}{2}\sum_{i=1}^{K}\log
_{2}(1+\frac{P_{i}^{\ast }\lambda _{i}^{2}}{N })$ (bits/symbol),
where $P_i^{\ast}$ are the waterfilling power allocations
$P_{i}^{\ast}=\left( \mu -\frac{N}{\lambda _{i}^{2}}\right) ^{+}$
in which $\mu$ is chosen to satisfy the power constraint
$\sum_{i=1}^{K}P_{i}^{\ast }=P$ and $P$ is the total power
allocated in one symbol time, and $\lambda_1$,...,$\lambda_{K}$,
are the singular values of the matrix $\mathbf{H}$. A more
detailed proof is presented in \cite{SOM}. According to the
formula of the capacity law, we obtain two important properties:
1) the capacity increases with the overlap factor $K$ because a
number of $K$ non-zero $\lambda_i$ exists in the meta-multiplexing
system \cite{SOM}, and each non-zero $\lambda_i$ corresponds to a
sub-channel that can support the transmission of one data stream;
2) the capacity depends on the chosen waveform, because different
waveforms generate significant different values for $\lambda_i$.

In communication practices, the energy per bit to noise power
spectral density ratio $E_b/N_0$ (also known as the ``SNR per
bit") is preferred to measure the condition of the communication
without taking bandwidth into account. Denote the sampling period
by $T_{s}$, and assume each data stream carries $\eta_{i}$ bits of
information; then, the capacity over $E_b/N_0$ is
$C=\frac{1}{2}\sum_{i=1}^{K}\log _{2}(1+\eta _{i}\frac{T_{s}}{T}
\frac{E^{\ast}_{b,i}}{N_0}\lambda _{i}^{2})$, where
$E^{\ast}_{b,i}=P_{i}^{\ast}T/\eta_i$ and $P_{i}^{\ast}$ is the
waterfilling power allocation \cite{SOM}. In practice, one simple
but common setting is to divide the power evenly for each stream,
$P_i=P/K$. In this case, the capacity becomes,
$C_{I}=\frac{1}{2}\sum_{i=1}^{K}\log_{2}(1+\eta
_{i}\frac{T_{s}}{T}\frac{E_{b}}{N_{0}}\lambda _{i}^{2})$.
Regarding to the capacity law, we first investigate the capacity
increasement with $K$. For simplicity, we choose a complex BPSK
signal $(+1,-1,+j,-j)$ as the input for the meta-multiplexing
system; thus, $\eta_i=2$. The capacity $C$ and $C_{I}$ are
calculated and shown in Fig. 3A. The Shannon capacity curve is a
special case of $C$ that $K=1$. To emphasize the linear
relationship between capacity and $E_b/N_0$, the capacity is also
displayed in a logarithmic scale. The numerical evaluation shows
that $C_{I}$ is quite close to the ideal capacity $C$.
Additionally, dividing the power evenly reduce the complexity and
the cost of implementation of the meta-multiplexing scheme; thus,
we use $C_{I}$ in the following discussion. To illustrate the
capacity under different waveforms, two Taylor waveforms, a
Gaussian waveform and a Hamming waveform, are evaluated and shown
in Fig. S5 in \cite{SOM}. It is obvious that the capacity of these
waveforms is much smaller than that of the rectangle wave because
(except for a few major ones) most of their singular values
attenuate much faster than singular values of the rectangle wave.
Therefore, the efficiency of the corresponding parallel channel is
smaller than the rectangle wave. However, in any circumstance,
when $K$ is sufficiently large, the capacity in a high SNR region
can still increase linearly with the SNR.

The structure of this communication is not that complex, why does
the community ignore it? One reason is that the meta-multiplexing
scheme completely breaks the Nyquist intersymbol interference
(ISI) criterion. The example depicted in Fig. 2A clearly shows the
intersymbol interference. Generally, when ISI occurs due to
various reasons such as the multipath propagation or bandlimit of
a channel, people tend to mitigate the effects
\cite{Tse_Viswanath_2005}. An alternative perspective is to
utilize intersymbol interference to help the communication system,
which dates back to the early 1970s \cite{Forney_1972}. However,
few later works have had significant success. One noticeable work
is the faster-than-Nyquist (FTN) signaling
\cite{Mazo_1975,Dasalukunte_etal_2014}, which intentionally builds
a controlled amount of ISI into the communication system. A small
amount of overlap on the symbols can carry more information with
an increased sample rate. However, this strategy limits the
overlap to less than half of a symbol duration.
The overlapped multiplexing principle proposed in
\cite{Daoben_2013}, further confirms that the overlapping between
adjacent symbols can be a beneficial constraint with coding gain,
but the overlap factor $K$ is limited to a small value. In this
work, we push the limit of intersymbol ``interference'' so that we
not only achieve impressive capacity when the overlap factor $K$
is sufficiently large but also disclose a capacity law that
reveals the reason why the capacity increases with $K$.

To design an entire communication system, conventional modules
like synchronization and channel estimation are required;
meanwhile, moderate modifications are needed to fit them into the
meta-multiplexing system. The conventional FEC code scheme can be
intergraded straightforwardly. One primary challenge lies in the
decoding process of the multiplexed signal: how to decode the
symbols, as they are heavily overlapped together. With regard to
the convolution structure of the meta-multiplexing, a
Viterbi-style algorithm is proposed. The trellis graph shows that
the unknown information bits are one-to-one mappings of the
trellis nodes, so decoding involves a dynamic search on the
trellis graph (see Fig. 2B for a example of the decoded path on
the trellis graph). However, the Viterbi-style algorithm can only
deal with small $K$, as the decoding complexity increases
exponentially with $K$. We further propose a Bayesian Monte Carlo
\cite{Book:Doucet_2001,Book:Liu_2001} based approximate decoding
method. The proposed algorithm uses simulation-based techniques to
simulate from a sequence of probability distributions for
sequential inference of the posterior distribution of unknown
information bits. The proposed algorithm provides a MAP decoding,
whereas the complexity is independent of $K$ \cite{SOM}. The
algorithm has also been designed for parallel computing to support
hardware implementation.


To simulate the transmission rate (bits per symbol) of the
meta-multiplexing over different overlap factor $K$, we choose the
complex BPSK signal as the input and the rectangle waveform as the
pulse shaping filter. We set a series value for the overlap factor
$K=[2, 4, 8, 10, 20, 30, 50,$ $60, 100, 200, 300, 450, 600, 900,
1200, 1800]$, and run the simulation study for $10^6$ bits. The
Bayesian Monte Carlo decoding can support decoding even when $K$
is even larger than $1000$. For each $K$, we evaluate the level of
$E_b/N_0$ that the system can gain the transmission rate of $2K$
bits/symbol with the bit error rate (BER) $\leq10^5$. The
simulated transmission rate in bits/symbol over $E_b/N_0$ is shown
in Fig. 3A. Obviously, the simulation results support that the
realized transmission rate is linearly proportional to the
$E_b/N_0$. The discrepancy between the realized transmission rate
and the theoretical capacity can be further reduced with the help
of a near-capacity forward error correcting (FEC) coding scheme,
such as turbo code \cite{Berrou_Glav_Thi_1993} and LDPC
\cite{Mackay_Neal_1996,Gallager_1963}. A twofold FEC coding
mechanism is applied in our meta-multiplexing system (see Fig. S1
in \cite{SOM}): the convolution structure of the meta-multiplexing
serves as the inner ``convolutional'' coding, which is a
noticeable advantage of our system. For outer FEC coding, we
choose a $3/5$ code rate LDPC from the DVB-S.2 standard. An
interleaving algorithm is inserted between the inner and outer
coding. Simulation studies in Fig. 3A illustrate that with the
help of LDPC coding, the realized transmission rate moves
approximately 4dB closer to the theoretical capacity curve.

The symbol, mentioned in previous discussion, actually represents
a two-dimensional resource containing both the time duration and
the frequency bandwidth. However, the dilemma is that for any
spectrum bandlimited signal, its waveform is not realizable, and
for any realizable waveform, its absolute bandwidth is infinite.
In engineering practices, various definitions of the bandwidth for
the realizable waveform have been proposed to describe the power
distribution of the signal \cite{SOM}. In the meta-multiplexing
system, one important property is that the multiplexed signal
follows the occupied bandwidth of the pulse shaping filter: the
convolution structure of the meta-multiplexing decides that the
bandwidth of the multiplexed signal is concentrated and generally
less than the bandwidth of the pulse shaping filter. Fig. S4
verifies the spectrum property of the meta-multiplexing signal
\cite{SOM}. Given the symbol's occupied bandwidth $B$, the
capacity of the meta-multiplexing can be expressed as
$\eta=\frac{1}{2BT}\sum_{i=1}^{K}\log _{2}(1+\frac{P_{i}\lambda
_{i}^{2}}{N })$ ( or $\frac{1}{2BT}\sum_{i=1}^{K}\log_{2}(1+\eta
_{i}\frac{T_{s}}{T}\frac{E_{b}}{N_{0}}\lambda _{i}^{2})$ for the
measurement by $E_b/N_0$) (bits/s/Hz) \cite{SOM}. The normalized
capacity (or spectral efficiency) is proportional to the capacity
with a coefficient $1/BT$. In a similar setting as discussed
before, we simulated the spectral efficiency $E_b/N_0$. The
resulting spectral efficiency (in Fig. 4A) shows that the curve
moves to the right compared with the capacity curve because that
the occupied bandwidth is always larger than $1/T$, and thus, the
coefficient $1/BT<1$.

According to the Nyquist-Shannon sampling theorem
\cite{Shannon_1948}, the processing bandwidth (the reciprocal of
the sampling rate) of a communication signal is at least two times
larger than the occupied bandwidth. In the meta-multiplexing, the
processing bandwidth becomes much larger than the occupied
bandwidth, particularly when $K$ is large (shown in Fig. S5 in
\cite{SOM}). A concern of this multiplexing scheme is how to
understand the ``problematic'' spectrum region that is outside of
the occupied bandwidth but inside of the processing bandwidth.
Here, we provide a convincing way to eliminate this concern: when
the ``problematic'' spectrum region is occupied by other
communication signals, either meta-multiplexing signals or
ordinary ones, the meta-multiplexing system can still work
properly. Fig. S6 in \cite{SOM} shows an example in which one
256-QAM signal exits in the processing bandwidth of a
meta-multiplexing signal. We define the bounded PSD at 35 dB as
the occupied bandwidth; then, the meta-multiplexing signal
occupies a bandwidth of 31.6 KHz, while the QAM occupies 750 KHz
approximately $75\%$ of the entire processing bandwidth of 1 MHz.
A joint decoding strategy is proposed to decode the signal (see
\cite{SOM} for details). The simulated spectral efficiency
requires approximately 2 dB higher $E_b/N_0$ than the one without
any neighboring signal in the processing bandwidth, but the linear
relation with the $E_b/N_0$ still holds (see the spectral
efficiency with spectrum sharing for different $K$ in Fig. 4A).
Consequently, we can measure the spectral efficiency of the
meta-multiplexing only on its occupied bandwidth, and then, the
resulting spectral efficiency is consistent linear increase with
the SNR, as we discussed previously. Moreover, the proposed scenario bring a smart strategy for the cognitive radio community\cite{Mitola_1999,Wang_Liu_2011}, in which the licensed primary users of the spectrum use the QAM, while the secondary users use the spectrum without interrupt and interaction with primary users.


Moreover, to verify that the meta-multiplexing works in real-world
physical channels, we implemented the entire communication system
on a standard verification system, the universal software radio
peripheral (USRP). The placement of the connection of the USRP
devices is shown in Fig. S7 in \cite{SOM}. All of the algorithms
were implemented on the onboard high-performance
field-programmable gate array (FPGA). We set $K=128$ and choose
the Taylor waveform with an attenuation level of -35dB for the
pulse shaping filter \cite{SOM}. The artificial delay $T/K$ is 1
millisecond. The meta-multiplexing signal is up-converted to radio
frequency at 2.4 Ghz. The spectrum of the channel signal is
measured by the spectrum analyzer and displayed in Fig. S8: the
measured occupied bandwidth of the channel signal is 24.48 KHz.
The noise generator add the Gaussian noise to the channel to
evaluate the BER performance. A pilot signal is utilized to
estimated the channel conditions for equalization before decoding.
We engineer the Bayesian Monte Carlo based approximate decoding
algorithm in a parallel style, which can fit the FPGA
implementation. Accordingly, all of the defects of a real physical
communication system, like analog-to-digital converter
(ADC)/digital-to-analog converter (DAC) quantization errors and
nonlinearity of the power amplifier, exist in this verification.
As shown in Fig. 4B, the dispersion between the simulation and
hardware implementation is caused by the effective number of bits
(ENOB) of ADC and DAC in the system. The BER of this hardware
implementation, which is highly consistent with the simulation
with ENOB=12, confirms that the proposed meta-multiplexing works
in practical communication systems. We utilized a 24.48 KHz
bandwidth but realized a reliable transmission
($\verb"BER"<10^{-5}$) of 2M bits/sec in the $E_b/N_0$ region of
45 dB, so the spectral efficiency is up to 81.7 bits/s/Hz, which
is significantly higher than the efficiency of any existing
communication system. Actually, such a high spectral efficiency is
generally infeasible for conventional communication methods like
high-order QAM modulations: the required ADC and DAC ENOB should
be larger than 40, and the required $E_b/N_0$ should be more than
230 dB, all of which are beyond the physical ability of
conventional communication systems.


The meta-multiplexing strategy can also be applied in other
domains, such as the frequency domain and joint time-frequency
domain, to create other novel multiplexing methods. Take
meta-multiplexing in frequency as an example: the superposition in
frequency would save bandwidth and benefit broadband communication
applications. The proposed communication framework is suitable for
different channel conditions. For example, in the multipath
Rayleigh fading channel, the random time dispersion of the channel
can be treated as part of the entire pulse response
$\textbf{h}(t)$, and blind deconvolution techniques can be
employed for the decoding when part of $\textbf{h}(t)$ is unknown.
As a general multiplexing technique offering significant capacity
enhancement, meta-multiplexing will bring new possibilities for
modern communications jointly with current communication
technologies.

\bibliography{scibib}

\begin{thebibliography}{10}
\bibitem{SOM}
Materials and methods are available as supporting material.
\end{thebibliography}

\bibliographystyle{science}


\begin{scilastnote}
\item We thank Prof. Daoben Li, Junbi Chen for their helpful discussions and kind suggestions, Dr. Changwei Lv, Chao Fang for their
assistance for the experimental apparatus.
\end{scilastnote}


\section*{{{\bf List of Figures}}}

\begin{figure}
\centerline{\includegraphics[width=14cm]{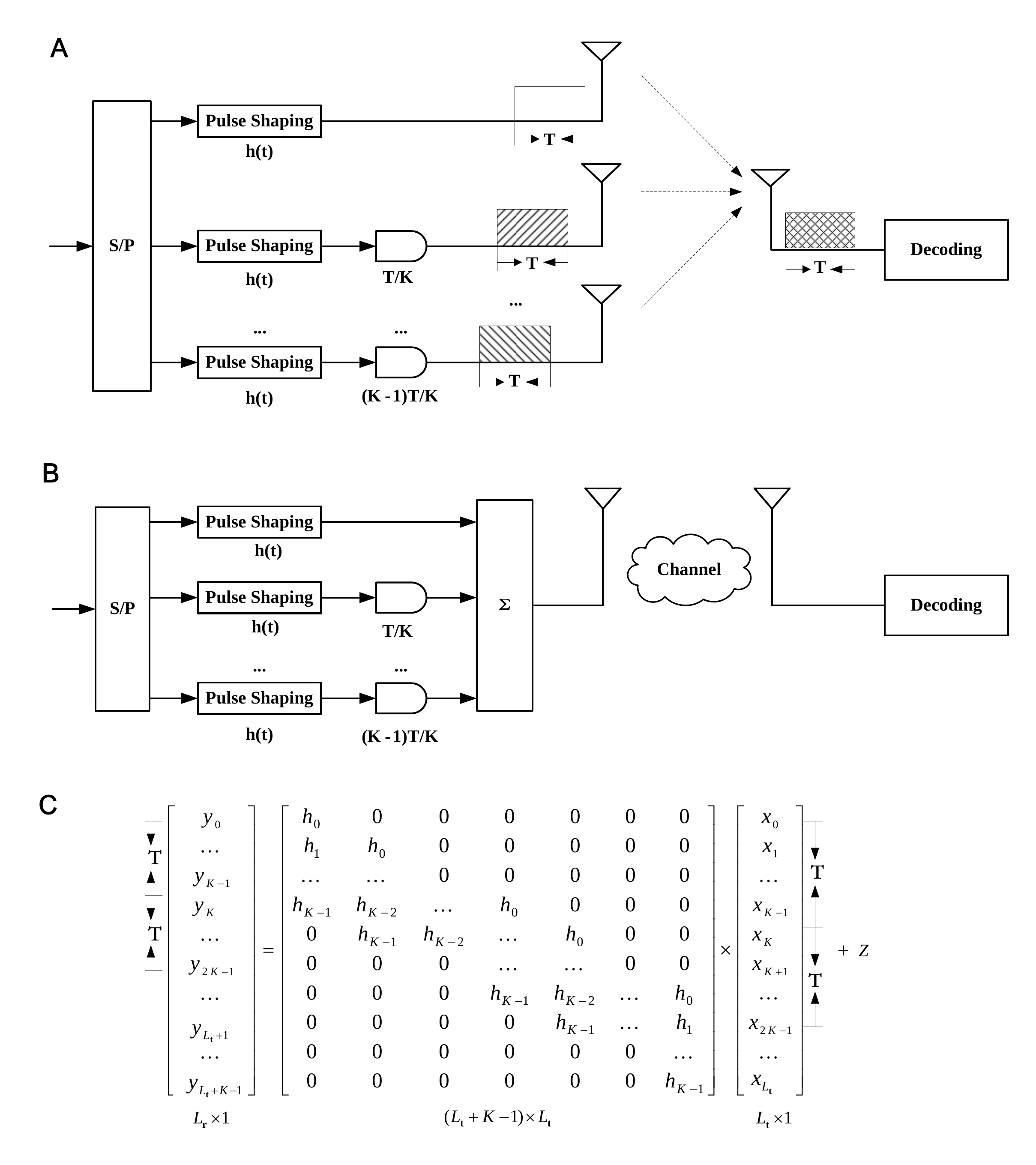}}
\caption{\small{The meta-multiplexing scheme. (\textbf{A}) In
meta-multiplexing, the information bits are paralleled to $K$
streams, and an artificial time delay is introduced for each
stream of the signal, then transmitted over a MISO system;
(\textbf{B}) The combination of the signal in multiple antennas
can be further simplified and processed digitally; the
meta-multiplexing then requires one single transmitter antenna.
(\textbf{C}) The system corresponds to a matrix expression with an
artificial channel matrix $H$.}} \label{fig1}
\end{figure}

\begin{figure}
\centerline{\includegraphics[width=14cm]{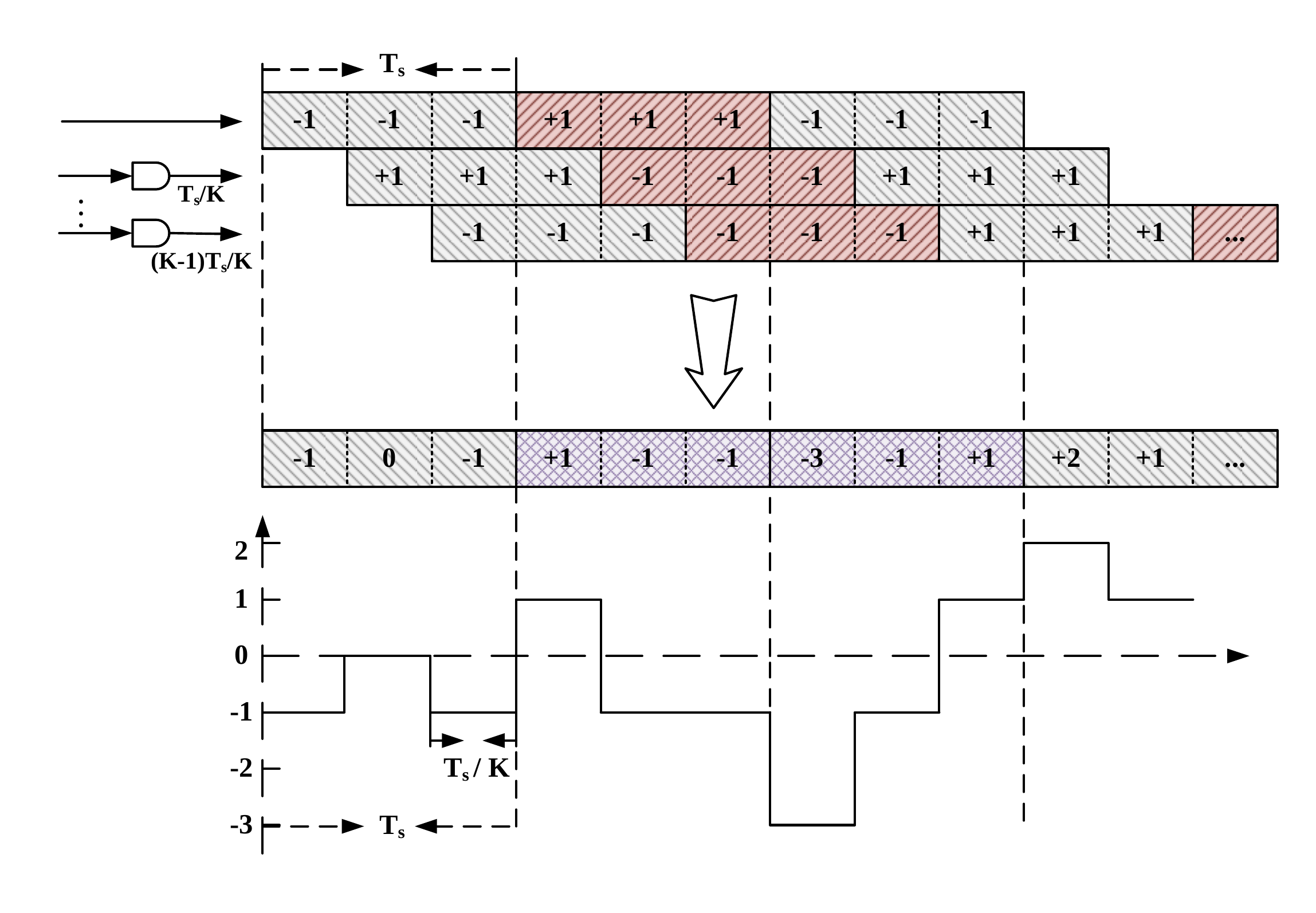}}
\centerline{\includegraphics[width=18cm]{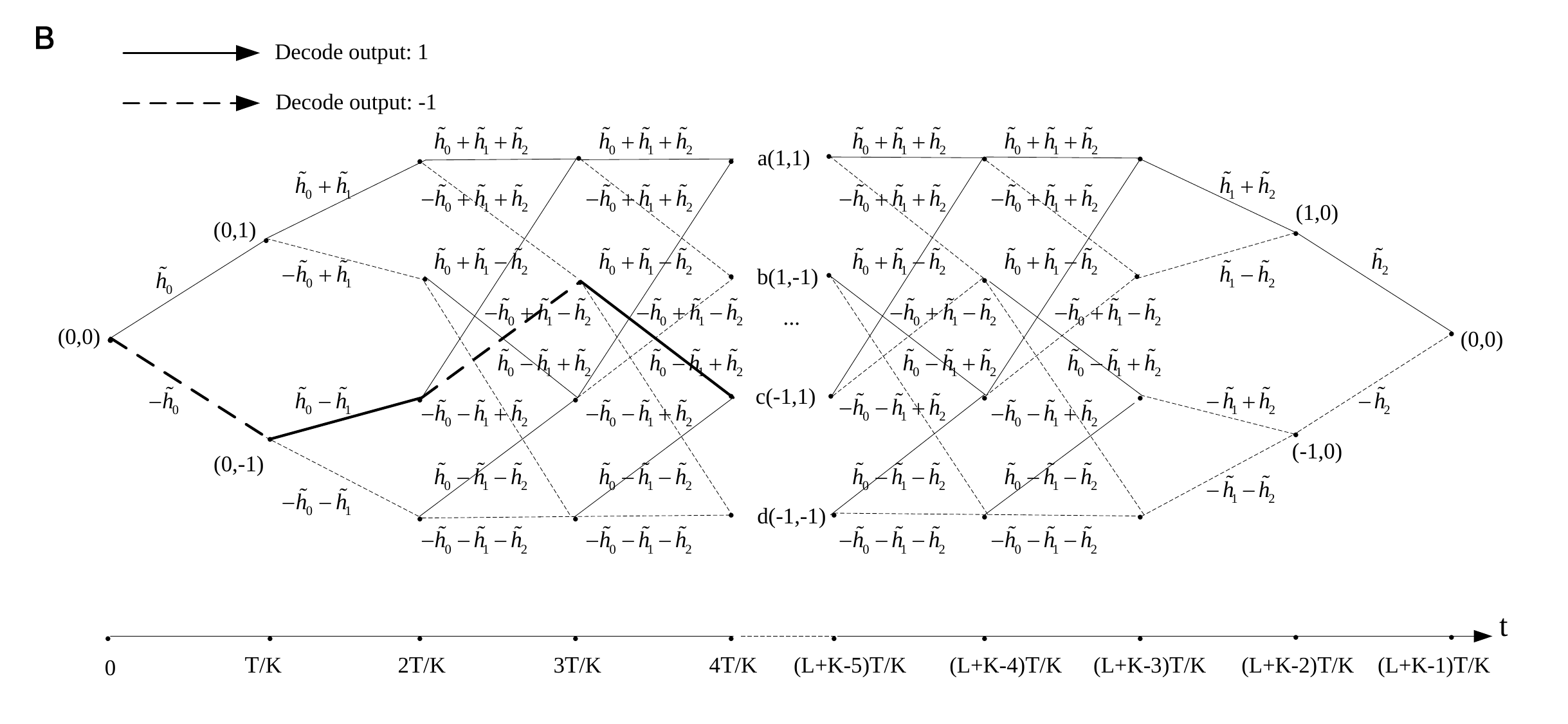}}
\caption{\small{(\textbf{A}) A simple case of the
meta-multiplexing scheme in which $K$=$3$ and $h(\cdot)$ is a
rectangle waveform. The modulated signal is a waveform with
maximally $K+1$ levels in each symbol time $T$. The trellis graph
(\textbf{B}) shows the idea of a Viterbi-style maximum likelihood
sequence detection method for decoding of the signal in
(\textbf{A}): in each time step, the survived path is found in
which the nodes have the minimum Euclidean distance with the
received signal.}}\label{fig2}
\end{figure}

\begin{figure}
\centerline{\includegraphics[width=9cm]{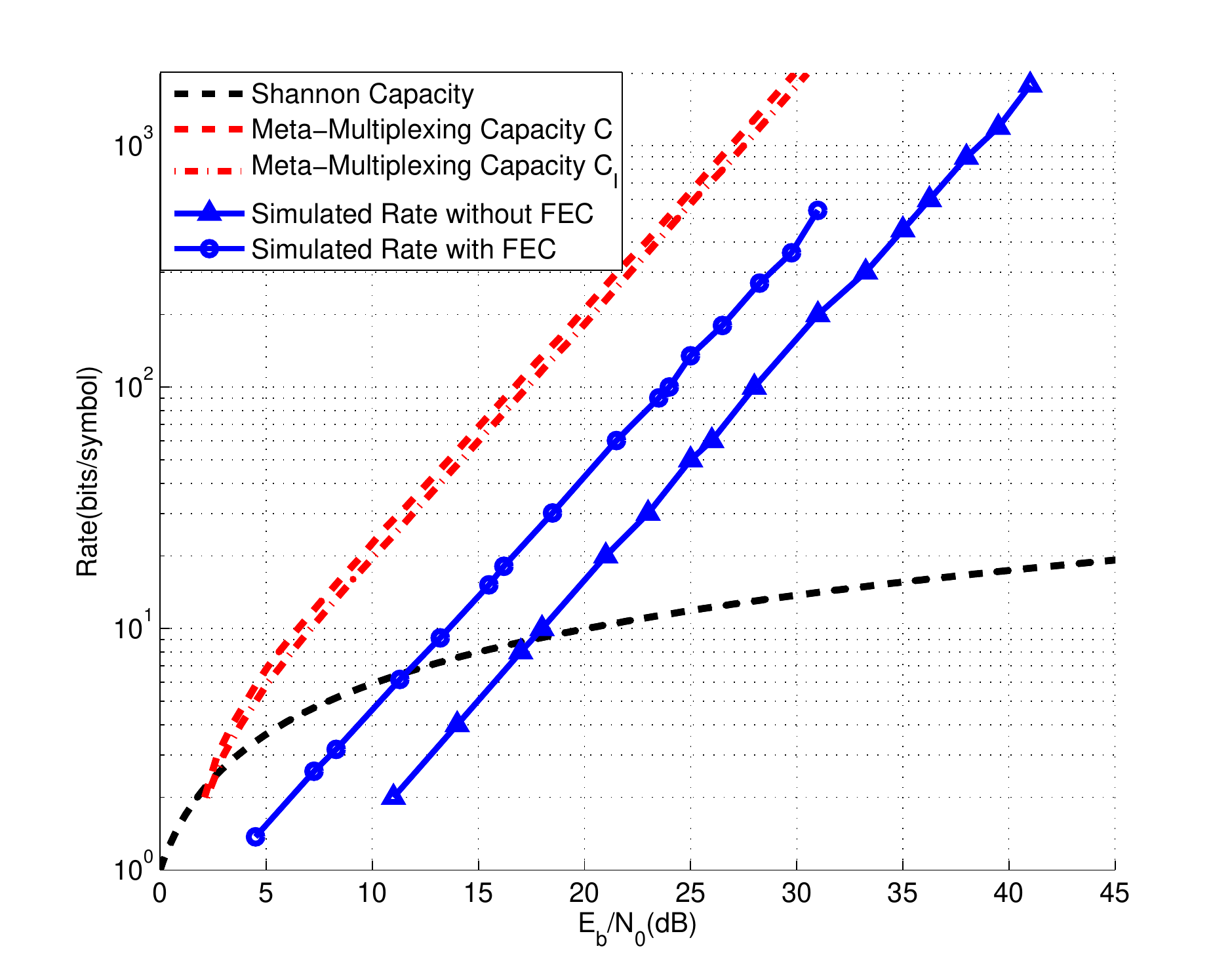}}
\centerline{\includegraphics[width=9cm]{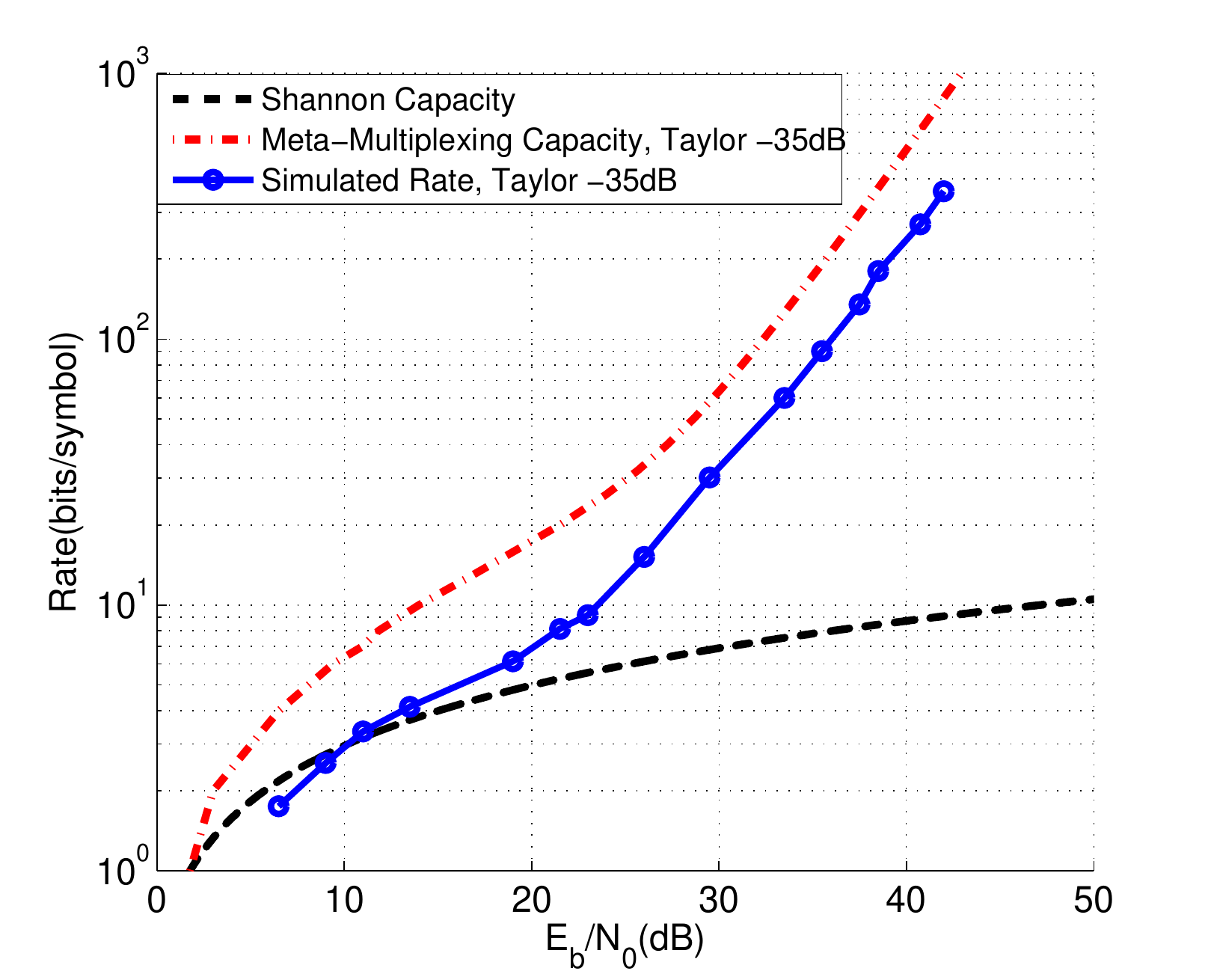}}
\caption{\small{(\textbf{A}) The capacity of meta-multiplexing
over different $K$ values, $C$ in dashed lines and $C_I$ in
dash-dot lines, show the linear relation between the capacity and
the $E_b/N_0$. The simulation of the realized transmission rate of
the meta-multiplexing system, solid lines with triangle, also
supports the linear relation. With a 3/5 LDPC coding, the
simulated transmission rate comes much closer to the theoretical
capacity curve; compared with (\textbf{A}), plots in (\textbf{B})
show the capacity and simulated transmission rate with the Taylor
waveform (with an attenuation level of -35 dB) and demonstrate
that the capacity of the meta-multiplexing scheme is highly
dependent on the designed waveforms. The linear relation between
the capacity and $E_b/N_0$ appears in a higher $E_b/N_0$
region.}}\label{fig3}
\end{figure}

\begin{figure}
\centerline{\includegraphics[width=9cm]{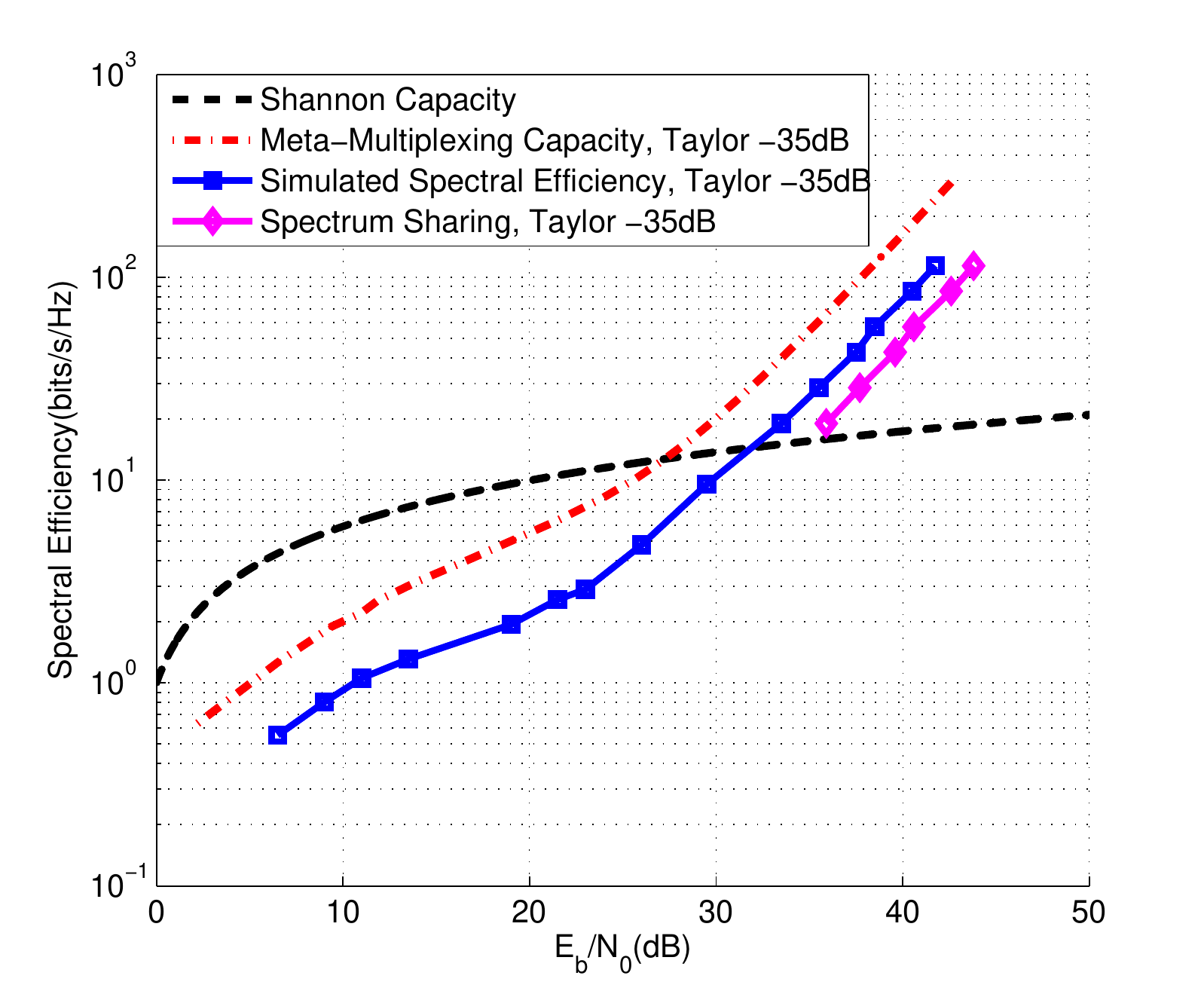}}
\centerline{\includegraphics[width=9cm]{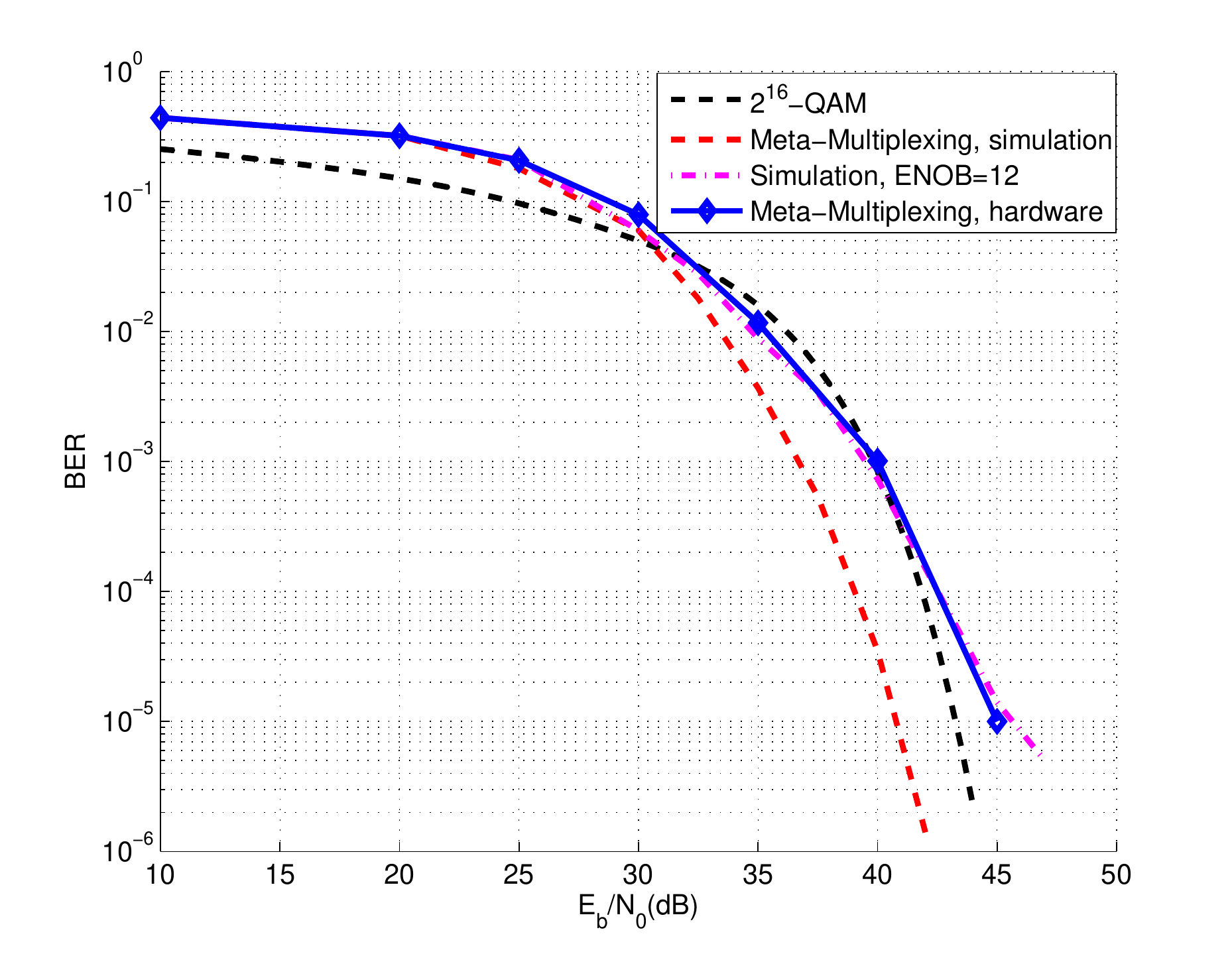}}
\caption{\small{(\textbf{A}) Simulation and compare study of the
normalized meta-multiplex capacity and the simulated spectral
efficiency of the system with the Talyor waveform; in addition,
the spectral efficiency with spectrum sharing with the QAM signal
is also drawn on the plot. Both the spectral efficiency and the
one with spectrum sharing is reasonably close to the theoretical
capacity curve, and both posses a linear increase region when
$E_b/N_0$ is sufficiently large; (\textbf{B}) Hardware
verification of a meta-multiplexing system in which $K=128$ and
$h(\cdot)$ is a Taylor waveform. The transmission rate is 256
bits/symbol, and the realized spectral efficiency is 81.7
bits/s/Hz. A comparison study with simulated spectral efficiency
shows that given the current hardware setting, the major influence
of the BER performance is the limitation of the ENOB of ADC and
DAC. Moreover, the BER of a high order $2^{16}$-QAM is presented.
According theoretical calculations, to achieve a spectral
efficiency of 81 bits/s/Hz, the QAM needs an extra 195 dB in the
SNR.}}\label{fig4}
\end{figure}

\clearpage
\newpage
\section*{Supporting Online Material}
\subsection*{The capacity of the parallel Gaussian channel
in the meta-multiplexing}

In the information theory, the concept of the parallel channel,
which refers to a set of non-interfering sub channels, has been
well accepted and leverages the capacity theory of MIMO systems
[1]. For meta-multiplexing, we have shown in Fig. 1 that the
system can be expressed as
\[
\mathbf{y} = \textbf{Hx} + \textbf{z}, \tag{S1}
\]
where vector \(\textbf{x}\in\textbf{C}^{L_{t}}\) denotes the
transmitted signal, vector \(\textbf{y}\in\textbf{C}^{L_{r}} \)
denotes the received signal, and \textbf{z} denotes the additive
white Gaussian noise that \(\textbf{z} \sim\) \(\textbf{N}(0,
N\textbf{I}_{L_{r}})\) at a sample time. \(L_t\) and \(L_r\)
denotes the length of transmitted and received signal, and
\textbf{C} denotes the complex field. The matrix \(\textbf{H}\in
\textbf{C}^{L_r \times L_t}\) represents the entire
meta-multiplexing process, in which each row is the impulse
response coefficient \textbf{h} shifted with a delay of \(1/K\)
symbol time (equal to a sample time). Due to the special structure
of \textbf{H}, it is easy to verify that the rank of matrix
\textbf{H} is \(L_t\).

To compute the capacity of this vector transmission model, we
decompose it into a set of parallel, independent Gaussian
sub-channels [1]. Apply the singular value decomposition,
\textbf{H=U\(\Lambda \textbf{V}^*\)} where \(\textbf{U}\in C^{L_r
\times L_r}\) and \(\textbf{V}\in C^{L_t \times L_t}\) are unitary
matrices. \(\Lambda\in\Re^{L_r\times L_t}\) is a rectangular
matrix, the diagonal elements of which are non-negative real
numbers and the off-diagonal elements of which are zero. The
diagonal elements of \(\Lambda\), \(\lambda_1 \geq \lambda_2 \geq
... \geq \lambda_{n_{min}}\), are the ordered singular values of
the matrix \textbf{H}. Denote \(\widetilde
\textbf{x}=\textbf{V}^*\mathbf{x}\), \(\widetilde
\textbf{y}=\textbf{U}^*\mathbf{y}\), and \(\widetilde
\textbf{z}=\textbf{U}^*\mathbf{z}\) and then, we obtain the
transformed expression
\[
\widetilde \textbf{y}=\Lambda \widetilde \textbf{x} + \widetilde
\mathbf{z} \tag{S2}
\]

Two properties hold [1]: \(\widetilde z \sim N(0,~N
\textbf{I}_{L_r})\) has the same distribution as \textbf{z}, due
to the Gaussian variable property; the transmission energy is
preserved, that \(\vert\vert\widetilde\textbf{x} \vert\vert ^2
=\vert\vert\textbf{x}\vert\vert ^2 \). Therefore, the
meta-multiplexing system (S1) has an equivalent representation as
the parallel Gaussian channel:
\[
\widetilde {y_i}=\lambda_i \widetilde {x_i} + \widetilde {z_i},
i=1,2,...n_{min} \tag{S3}
\]
The equivalence is depicted in Fig. S3.

We consider a simple case first, in which the number of
information bits \(L_t = K\), then \(L_r = 2K-1\), and
\(n_{min}=K\) . Now, the rank of \textbf{H} is \(K\); thus, the
number of nonzero singular values is also \(K\). The capacity of
the parallel channel (S3) is the maximum of mutual information:

\[
C= \mathop {}_{\sum E[\widetilde x_i^2] \leq P}^{max}
I(\widetilde x_1,\widetilde x_2,...\widetilde x_K,; \widetilde
y_1, \widetilde y_2,...\widetilde y_K), \tag{S4}
\]
where \(P\) is the power constraint. Firstly, we need to prove
that the capacity of the meta-multiplexing system (S1) has the
same mutual information of the parallel channel (S3). Regards to
the equation (S2), when \(L_t = K \), only \(n_{min} = K \)
channel exists which means \(\lambda_i = 0\) for \(i \textgreater
K \). Thus, \(\widetilde y_i = \widetilde z_i \)  for \(i
\textgreater  K \). By utilizing the chain rule for the entropy
[2], we obtain that

\[
  \begin{split}
&h(\widetilde y_1,\widetilde y_2,...,\widetilde y_K,\widetilde y_{K+1},...,\widetilde y_{L_r}) \\
&=h(\widetilde y_1,\widetilde y_2,...,\widetilde y_K) + h(\widetilde y_{K+1},...,\widetilde y_{L_r} \vert \widetilde y_1,\widetilde y_2,...,\widetilde y_K) \\
&=h(\widetilde y_1,\widetilde y_2,...,\widetilde y_K)+h(\widetilde y_{K+1},...,\widetilde y_{L_r}) \\
&=h(\widetilde y_1,\widetilde y_2,...,\widetilde y_K)+h(\widetilde z_{K+1},...,\widetilde z_{L_r}), \\
   \end{split}
   \tag{S5}
\]
For the noise \textbf{z}, the chain rule also holds:
\[
h(\widetilde z_1,\widetilde z_2,...\widetilde z_K,\widetilde
z_{K+1},...,\widetilde z_{L_r}) =h(\widetilde z_1,\widetilde
z_2,...,\widetilde z_K)+h(\widetilde z_{K+1},...,\widetilde
z_{L_r}),   \tag{S6}
\]
A theorem about the entropy of a vector variable with matrix
operation is also needed here (chapter 8 in [2]):
\[
h(\widetilde \textbf{y}) = h(\textbf{U}^* \textbf{y}) =
h(\textbf{y}) + \log \vert \det(\textbf{U}^*) \vert =
h(\textbf{y}), \tag{S7}
\]
where \(\log \vert \det(\textbf{U}^*) \vert \) equals to 0, as
\(\textbf{U}^*\) is a unity matrix. Similarly, we have
\(h(\widetilde \textbf{z}) = h(\textbf{z})\). Given equations
(S5), (S6), and (S7), we can obtain that
\[  \begin{split}
&I(x_1, x_2, ..., x_K; y_1, y_2,...,y_{L_r}) \\
&=h(y_1, y_2, ..., y_{L_r}) - h(y_1, y_2, ..., y_{L_r} \vert x_1, x_2, ..., x_K) \\
&=h(y_1, y_2, ..., y_{L_r}) - h(z_1, z_2, ..., z_{L_r} \vert x_1, x_2, ..., x_K) \\
&=h(y_1, y_2, ..., y_{L_r})-h(z_1, z_2, ..., z_{L_r})\\
&=h(\widetilde y_1, \widetilde y_2, ..., \widetilde y_{L_r})-h(\widetilde z_1, \widetilde z_2, ..., \widetilde z_{L_r})\\
&=[h(\widetilde y_1, \widetilde y_2, ..., \widetilde y_K) + h(\widetilde z_{K+1},...,\widetilde z_{L_r})]-\\
&~~~~~[h(\widetilde z_1, \widetilde z_2, ..., \widetilde z_K) + h(\widetilde z_{K+1},...,\widetilde z_{L_r})] \\
&=h(\widetilde y_1, \widetilde y_2, ..., \widetilde y_K)-h(\widetilde z_1, \widetilde z_2, ..., \widetilde z_K)\\
&=h(\widetilde y_1, \widetilde y_2, ..., \widetilde y_K)-h(\widetilde z_1, \widetilde z_2, ..., \widetilde z_K \vert \widetilde x_1, \widetilde x_2, ..., \widetilde x_K)\\
&=h(\widetilde y_1, \widetilde y_2, ..., \widetilde y_K)-h(\widetilde y_1, \widetilde y_2, ..., \widetilde y_K \vert \widetilde x_1, \widetilde x_2, ..., \widetilde x_K)\\
&=I(\widetilde x_1, \widetilde x_2, ..., \widetilde x_K ;\widetilde y_1, \widetilde y_2, ..., \widetilde y_K) \\
   \end{split}
 \tag{S8}
\]

Now \(I(x_1, x_2, ..., x_K; y_1, y_2,...,y_{L_r}) = I(\widetilde
x_1, \widetilde x_2, ..., \widetilde x_K ;\widetilde y_1,
\widetilde y_2, ..., \widetilde y_K) \) holds, then we need to
find the upper bound of the mutual information. A well known
corollary in [2] that \(h(a_1, a_2,...,a_L) \leq \sum_{i=1}^{L}
h(a_i)\) with equality iff (if and only if) \(a_1, a_2,...,a_L\)
are independent is needed. Thus,
\[
h(\widetilde y_1,\widetilde y_2,...\widetilde y_K) \leq
\sum_{i=1}^{K} h(\widetilde y_i), \tag {S9}
\]

The Gaussian channel noise z = \(z_1, z_2 ,...,z_L \) are
independent and identically distributed. Here, we prove that
\(\widetilde \textbf{z} = \widetilde z_1, \widetilde z_2
,...,\widetilde z_L \)  are also independent: according to the
covariance matrix, \(Cov(\widetilde \textbf{z}) = Cov(\textbf{U}^*
\textbf{z})=\textbf{U}^*Cov(\textbf{z})\textbf{U}=N\textbf{U}^*\textbf{U}=N\textbf{I}_{L_r}
\) , \(Cov(\widetilde z_i, \widetilde z_j) =0 \) for \(i \neq j\)
; then, due to the property of Gaussian random variables that the
uncorrelation is equivalent to independence, so \(\widetilde
\textbf{z} = \widetilde z_1, \widetilde z_2,... \widetilde z_L \)
are independent [3]. Therefore, the equality holds,
\[
h(\widetilde z_1, \widetilde z_2, ..., \widetilde z_K) =
\sum_{i=1}^{K} h(\widetilde z_i) \tag {S10}
\]

To derive the closed-form expression of the capacity for Gaussian
channels, we need another theorem about entropy that given the
random vector \(\textbf{a} \in \Re ^n \) with zero mean and
covariance \(\sum = E[\textbf{aa}^T]\), then \(h(\textbf{a}) \leq
\frac{1}{2} \log(2 \pi e)^n \vert \sum \vert \) with equality iff
\(\textbf{a} \sim \textbf{N}(0,\sum)\) [2]. Denote the allocated
power to each \(x_i \) by \(P_i \) and the noise power by \(N\) .
Thus, by definition, \(P_i =E[x_i^2]\), \(N =E[z_i^2]\)=
\(E[\widetilde z_i^2]\). Moreover, because \(\widetilde y_i=
\lambda_i \widetilde x_i + \widetilde z_i \) and because
\(\widetilde x_i\) and \(\widetilde z_i\) are independent, the
average power \(E[\widetilde y_i^2]=\lambda_i^2 P_i +N \). Thus,
\[
h(\widetilde y_i) \leq \frac{1}{2} \log[2 \pi e(\lambda_i^2 P_i +
N)], \tag {S11}
\]
and
\[
h(\widetilde z_i)=\frac{1}{2} \log[2 \pi e N]. \tag {S12}
\]

Finally, given (S8)-(S12), the mutual information is bounded as
follows:

\[
  \begin{split}
&I(x_1, x_2, ..., x_K; y_1, y_2,...,y_{L_r}) \\
&=I(\widetilde x_1, \widetilde x_2, ..., \widetilde x_K; \widetilde y_1, \widetilde y_2,...,\widetilde y_K) \\
&=h(\widetilde y_1, \widetilde y_2, ..., \widetilde y_K)-h(\widetilde y_1, \widetilde y_2, ..., \widetilde y_K | \widetilde x_1, \widetilde x_2,...,\widetilde x_K)\\
&=h(\widetilde y_1, \widetilde y_2, ..., \widetilde y_K)-h(\widetilde z_1, \widetilde z_2, ..., \widetilde z_K | \widetilde x_1, \widetilde x_2,...,\widetilde x_K)\\
&=h(\widetilde y_1, \widetilde y_2, ..., \widetilde y_K)-h(\widetilde z_1, \widetilde z_2, ..., \widetilde z_K)\\
&=h(\widetilde y_1, \widetilde y_2, ..., \widetilde y_K)-\sum_{i=1}^{K} h(\widetilde z_i) \\
&\leq \sum_{i=1}^{K} h(\widetilde y_i) - \sum_{i=1}^{K} h(\widetilde z_i)\\
&=\sum_{i=1}^{K} h(\widetilde y_i) - \frac{1}{2} \sum_{i=1}^{K} \log[2 \pi e N] \\
&\leq \frac{1}{2} \sum_{i=1}^{K} \log[ 2 \pi e(P_i \lambda_i^2 + N)] - \frac{1}{2} \sum_{i=1}^{K} \log [2 \pi e N] \\
&= \frac{1}{2} \sum_{i=1}^{K} \log(1+\frac{P_i \lambda_i^2}{N})  \\
&\leq \frac{1}{2} \sum_{i=1}^{K} \log(1+\frac{P_i^*
\lambda_i^2}{N})
   \end{split}
\tag {S13}
\]
where \(P_1^* \), \(P_2^* \), ...\(P_K^* \) are the waterfilling
power allocations [1,2]:
\[
P_i^* = (\mu - \frac{N}{\lambda_i^2})^+ \tag {S14}
\]
where \(\mu\) is chosen to satisfy the total power constraint
\(\sum_{i=1}^{K} P_i^* = P \), and \(P\) is the total power
allocated in one symbol time.

Moreover, note that when \(L_t =K \), \(L_r = 2K-1 \)  samples are
transmitted in the channel, while \(K\) samples form a symbol;
therefore, we actually transmit 1+\((K-1)/K\) symbol. With respect
to the factor, the capacity is refined as
\[
C=\frac{1}{2} \frac{1}{1 +\frac{K-1}{K}}
\sum_{i=1}^{K}\log(1+\frac{P_i^* \lambda_i^2}{N}) (\text{bits per
symbol}) \tag {S15}
\]

For the case when \(L_t \) is larger than \(K\), it is easy to
obtain that
\[
C=\frac{1}{2} \frac{\frac{L_t}{K}}{\frac{L_t}{K} +\frac{K-1}{K}}
\sum_{i=1}^{K}\log(1+\frac{P_i^* \lambda_i^2}{N}) (\text{bits per
symbol}) \tag {S16}
\]

Thus, when \(L_t \) is sufficiently large, we can conclude that
the capacity of the meta-multiplexing system is
\[
C= \frac{1}{2} \sum_{i=1}^{K} \log(1+\frac{P_i^* \lambda_i^2}{N})
(\text{bits per symbol}), \text{as} ~ L_t \rightarrow {\infty}
\tag {S17}
\]

\subsection*{Derivation of the capacity over \(E_b/N_0\)}
We utilize the energy per bit to noise power spectral density
ratio \(E_b/N_0\) to measure the condition of the communication
channel. For each stream, denote the number of information bits
per symbol by \(\eta_i \), which might be influenced by the size
of the modulation alphabet or the code rate of an error-control
code in each stream. By definition, \(P_i^* = \frac{\eta_i
E_{b,i}^*}{T}\), in which \(E_{b,i}^*\) is the bit energy in each
sub-channel. For a complex communication signal, we have \(N=W_n
N_0=N_0/T_s\), where \(W_n\) is the noise bandwidth, \(N_0\) is
the noise power spectral density, and \(T_s\) is the sampling
period. According to the previous discussion, when each stream has
the same power, the capacity becomes

\[
   \begin{split}
C&= \frac{1}{2} \sum_{i=1}^{K} \log(1+\frac{P_i^* \lambda_i^2}{N})  \\
&=\frac{1}{2} \sum_{i=1}^{K} \log(1+\eta_i \frac{T_s}{T} \frac
{E_{b,i}^*}{N_0} \lambda_i^2).
   \end{split}
\tag {S18}
\]
In practices, one simple setting is that the power for each stream
is equally divided, thus \(P_i = \frac {P}{K}\).
\[
   \begin{split}
C_I &= \frac{1}{2} \sum_{i=1}^{K} \log(1+\frac{P_i \lambda_i^2}{N})  \\
&=\frac{1}{2} \sum_{i=1}^{K} \log(1+\eta_i \frac{T_s}{T} \frac
{E_b}{N_0} \lambda_i^2).
   \end{split}
\tag {S19}
\]

\subsection*{Evaluation of the capacity for different K}
We take\((+1, -1, +j, -j)\)as the input of each stream of the
meta-multiplexing, and the information carried by each symbol of
each stream is \(\eta_i =2 \) . Given the overlap factor \(K\),
then the meta-multiplexing system consists of \(K\) parallel
sub-channels and the total information carried by a symbol is
\(2K\). By using equation (S18), we can then evaluate the required
\(E_b/N_0\)to achieve this capacity \(C=2K\) (or \(C_I = 2k) \).
Consequently, we can produce the curve of the capacity over
different \(K\).

\subsection*{Definition of bandwidths}
It is common knowledge that for strictly bandlimited signals, the
waveforms are not realizable because the time duration is
infinite, whereas for realizable waveforms, the absolute bandwidth
is infinite. Therefore, in practices, the bandwidth for the
realizable waveform has been proposed to describe the power
distribution of the communication signal. In this study, we
consider two common definitions of the occupied bandwidth due to
different considerations [4]: 1) Bounded power spectral density
(Bounded PSD). This criterion states that everywhere outside of
the occupied band, the spectrum of the signal  \(H(f)\) must have
fallen to a specified level below the level at the band center.
Typical attenuation levels used in this study are 35 dB and 50 dB;
2) Fractional power containment bandwidth (FPCB). This bandwidth
criterion states the percent of the signal power inside the
occupied bandwidth. It has been adopted by the Federal
Communications Commission. Typical percentages commonly used are
99\%. Waveforms and their bandwidths used in this paper are listed
as follows:

\textbf{h}$_1(t)$: Rectangular waveform with time duration T. The
fall-off rate of the sidelobe of the rectangular wave is very low,
so its occupied bandwidth is much larger than \(1/T\): By
numerical calculation, the 35dB bounded PSD bandwidth is
\(B_{BPSD}=49.30/T\), the 50dB bounded PSD bandwidth cannot be
achieved within \(200/T\) , and the 99\% FPCB bandwidths are
\(18.28/T \) respectively. Rectangular wave is only employed to
demonstrate the idea of the meta-multiplexing and the decoding
algorithm, and is not suitable for real communication systems
because its occupied bandwidth is unacceptable in practice;

\textbf{h}$_2(t)$: Taylor waveform with time duration  and the
sidelobe level at 35dB, the 35dB bounded PSD bandwidth is
\(B_{BPSD}=3.16/T\) (see Fig. S4), and the 99\% fractional power
FPCB bandwidths is \(B_{FPCB}=2.35/T\);

\textbf{h}$_3(t)$: Taylor waveform with time duration  and the
sidelobe level at 50dB, the 50dB bounded PSD bandwidth is
\(B_{BPSD}=3.90/T\)(see Fig. S4), and the 99\% fractional power
FPCB bandwidths is \(B_{FPCB}=2.74/T\).

\subsection*{The normalized capacity}
Besides, generally the communication signal is sampled at a higher
frequency than the occupied bandwidth. When a Fourier transform
applied, the signal is distributed on a much wider bandwidth,
which is named processing bandwidth, while the bandwidth in which
most of the signal energy are concentrated is defined as the
occupied bandwidths. Commonly, to measure the spectral efficiency
of a communication system, we use the occupied bandwidths rather
than the processing bandwidth. Given the occupied bandwidth B and
time duration T of a meta-multiplexing symbol, the limit on
spectral efficiency of this communication system, measured by
bits/s/Hz, is

\[
\eta= \frac{1}{2BT} \sum_{i=1}^{M} \log(1+\frac{P_i
{\lambda}_i^2}{N})(\text{bits/s/Hz}), \tag {S20}
\]

\[
~~~~~~~~~~~~~~~~=\frac{1}{2BT} \sum_{i=1}^{M} \log(1+\eta_i
\frac{T_s}{T} \frac {E_b}{N_0} \lambda_i^2) (\text{bits/s/Hz}).
\tag {S21}
\]

\subsection*{Bayesian Monte Carlo based approximate decoding}
First, the meta-multiplexing system must be reformed as a
state-space model by the following two processes: the state
equation, which acutely represents the convolution process of
information bits passing through the meta-multiplexing system,
\(s_t = \sum_{i=0} ^ {K-1} h_k x_{t-k}\)  ($t =1,2,...$), in which
\(x_t\) represents the information bits and \textbf{h}
$\doteq[h_0,...,h_{k-1}]$ represents the impulse response
coefficients of the entire communication system, including the
pulse shaping filter, channel impulse response, matched filter,
and etc. The second process is the observation equation, which is
\(y_t\) = \(s_t\) + \(z_t\), in which \(z_t\) is the additive
channel noise. Given the state-space model, Bayesian sequential
inference can provide the estimation of the posterior of
\(p(x_t\vert y_{1:t})\), which further leads to the maximum a
posterior (MAP ) estimation of \(x_t\) . Sequential Monte Carlo
(SMC), also known as particle filer, is a class of importance
sampling and resampling techniques designed to solve the Bayesian
inference of \(p(x_t\vert y_{1:t})\)[5, 6, 7]. Here, we present
one typical SMC method, the sequential importance resampling (SIR)
method [5]. We need to design the proposal distribution \(\pi(x_t
\vert x_{1:t-1} , y_{1:t})\), a probability density function from
which it is easy to sample, whereas in our case, we can simply use
the prior distribution \(Ber (1/2)\), as \(x_t\) is subject to a
Bernoulli distribution and independent of \(x_{1:t-1}\).With a
simple initiation, the SIR can be summarized as follows:

At iteration \(t \geq 1\), for all \(i=1,...,M\):

\textbf{Step 1}: Sample \(x_t^{(i)}\sim Ber (1/2)\)   and set
\(x_{1:t}^{(i)} = [x_{1:t-1} ^{(i)}, x_t^{(i)}] \);

\textbf{Step2}: Compute the weights \(\widetilde \omega_t^{(i)}\)
\(\sim\) \(\widetilde \omega_{t-1}^{(i)} (p(y_t \vert x_t^{(i)})\)
, where \(\widetilde \omega_t^{(i)}\)  is normalized as
\(\sum_{i=1}^{M} \)\(\widetilde \omega_t^{(i)}\) =1;

\textbf{Step3}: If the effective number of particles
$N_{\texttt{eff}}=\frac{1}{\sum_{i=1}^{M}
(\widetilde\omega_t^{(i)})^2 }$ is less than a threshold, then
perform resampling \(x_{1:t}^{(i)}\) \(\sim \sum_{i=1}^{M}
\)\(\widetilde \omega_t^{(i)}\) \(\delta (x_{1:t} ^{(i)})\).

According to the above process, it is easy to notice that the SIR
algorithm is suitable for parallel computing, which significantly
benefits the hardware implementation [8]. Moreover, the
computational complexity of the SIR algorithm depends only on the
number of particles , so it is suitable for cases when the overlap
factor  is large in the meta-multiplexing. In the practice of
communication systems, the coefficient \textbf{h} may be unknown
or with uncertain. In this case, we can either utilize some pilot
signal to estimate the coefficients or extend the SMC to jointly
estimate the state \(x_{1:t}\)  and the coefficient \textbf{h}
[9].

\subsection*{Joint decoding strategy}
The joint decoding strategy attempts to demodulate all of the
signals in the processing bandwidth of the meta-multiplexing.
Generally, we assume that all other communication signals in the
processing bandwidth are cooperative and we can demodulate the
signal when the SNR is high enough. Without a loss of generality,
we assume only one QAM modulated signal coexists with the
meta-multiplexing signal in the processing bandwidth. Then, the
decoding can be processed as follows:

\textbf{Step 1}: Use the coherent demodulation technique to move
the QAM signal to baseband, and then filter out the signal outside
of the occupied bandwidth of the QAM signal;

\textbf{Step 2}: Demodulate the QAM signal by a series of
processes: matched filtering, down sampling, and detection;

\textbf{Step 3}: In the Bayesian approximate decoding process for
the meta-multiplexing, use the demodulated QAM bits and the
proposed bits \(x_t^{(i)}\) for meta-multiplexing to generate the
multiplexing signal \(s_t^{(i)}\) (where \(i\) is the index of
each particle);

\textbf{Step 4}: Measure the distance between the observation and
each particle by the likelihood function \(p (y_t \vert
s_t^{(i)})\), and perform the reweighting and resampling in the
SIR algorithm;

\textbf{Step 5}: Obtain the MAP estimate of the meta-multiplexing
based on the particles \(\{{x_{1:t}^{(i)}}\}_{i=1}^M\).

\subsection*{Hardware verification aperture and settings}
The USRP devices become a common tool to verify communication
methods and prototypes. In this study, we utilize a pair of USRP
devices with programmable FPGA modules and configurable
radio-frequency modules. The placement of the hardware
connectivity is shown in Fig. S7: the transmitter (USRP Tx) and
receiver (USRP Rx) are connected by a radio frequency (RF) cable,
and a combiner is inserted to connect the noise generator and
spectrum analyzer. Fig. S8 shows the spectrum of the
meta-multiplexing signal measured by the spectrum analyzer. All of
the communication algorithms are implemented on the FPGA, so the
system can run stand-alone. Moreover, both the transmitter and
receiver USRP devices are connected to a computer to verify the
reliability of the communication system and calculate the bit
error rate (BER).

\newpage

\begin{figure}[htbp]
\centerline{\includegraphics[width=10cm]{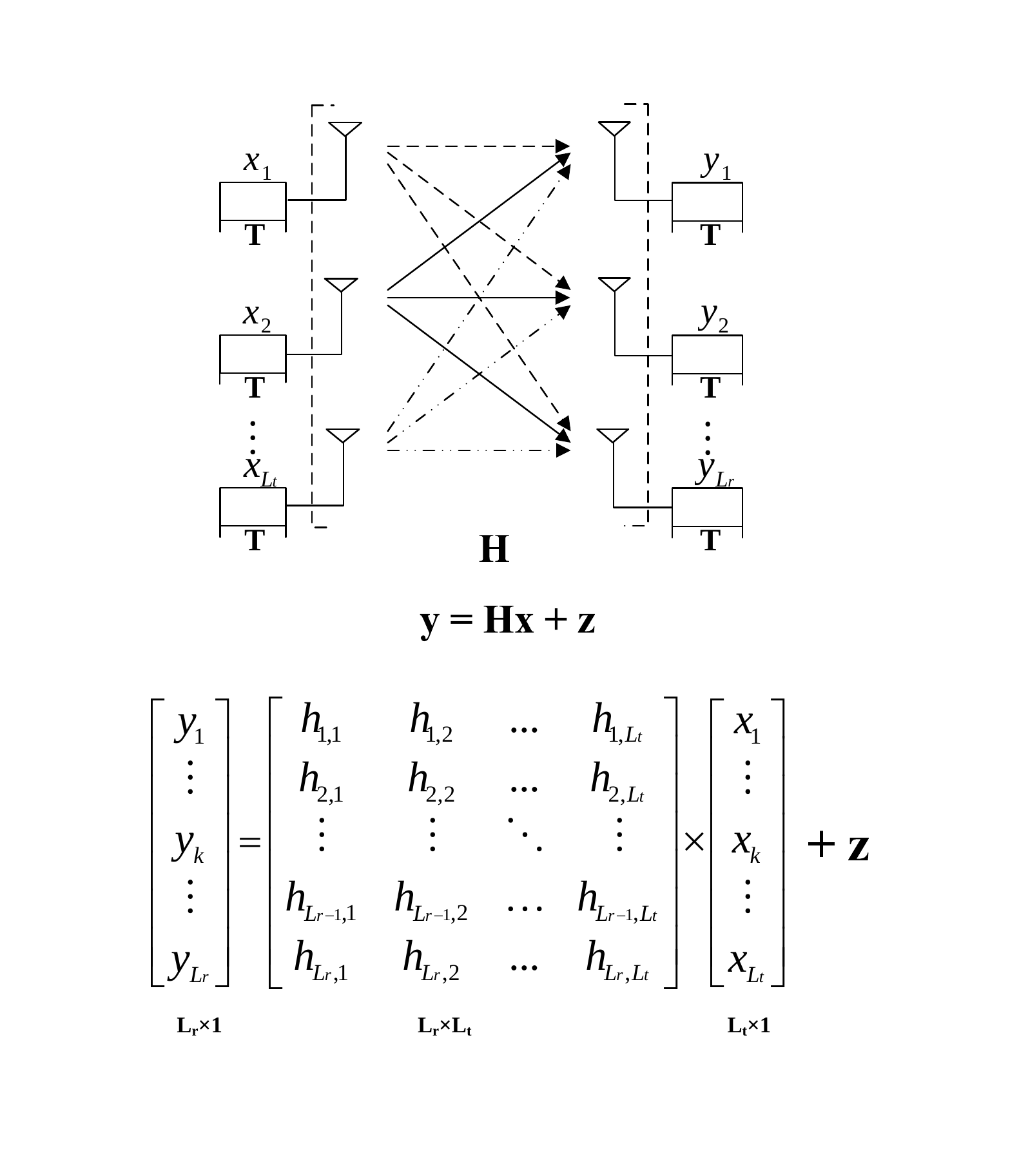}}
\small{Fig. S1. Schematic of a MIMO system: the information bits
are formatted into multiple streams, transmitted through
transmission antennas and a multipath channel, and received by
receiver antenna. The MIMO system can be expressed as , where the
matrix represents the physical paths between the transmitter and
the receiver [1].} \label{default}
\end{figure}

\begin{figure}[htbp]
\centerline{\includegraphics[width=14cm]{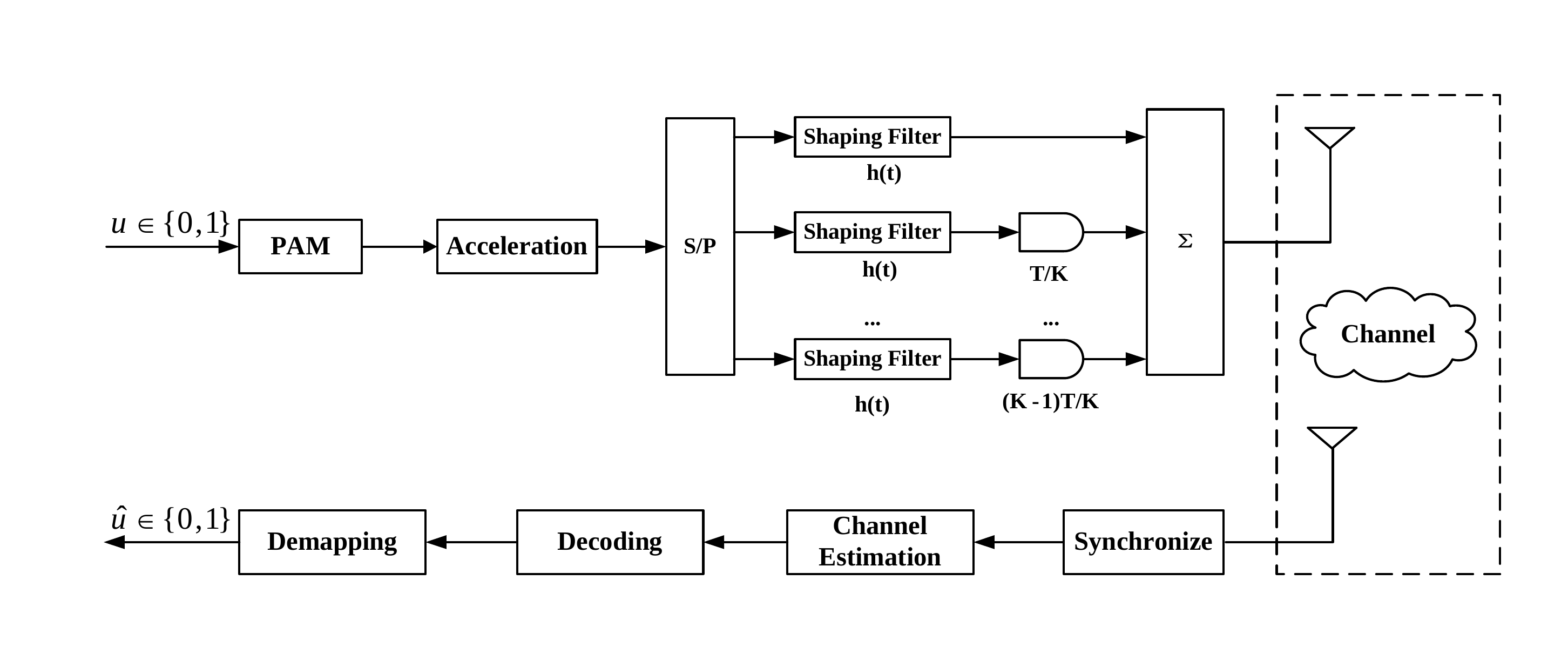}}
\small{Fig. S2. The framework of a meta-multiplexing communication
system. Conventional communication modules like FEC coding,
synchronization and channel estimation requires only moderate
modification to fit into the meta-multiplexing system.}
\label{default}
\end{figure}

\begin{figure}[htbp]
\centerline{\includegraphics[width=12cm]{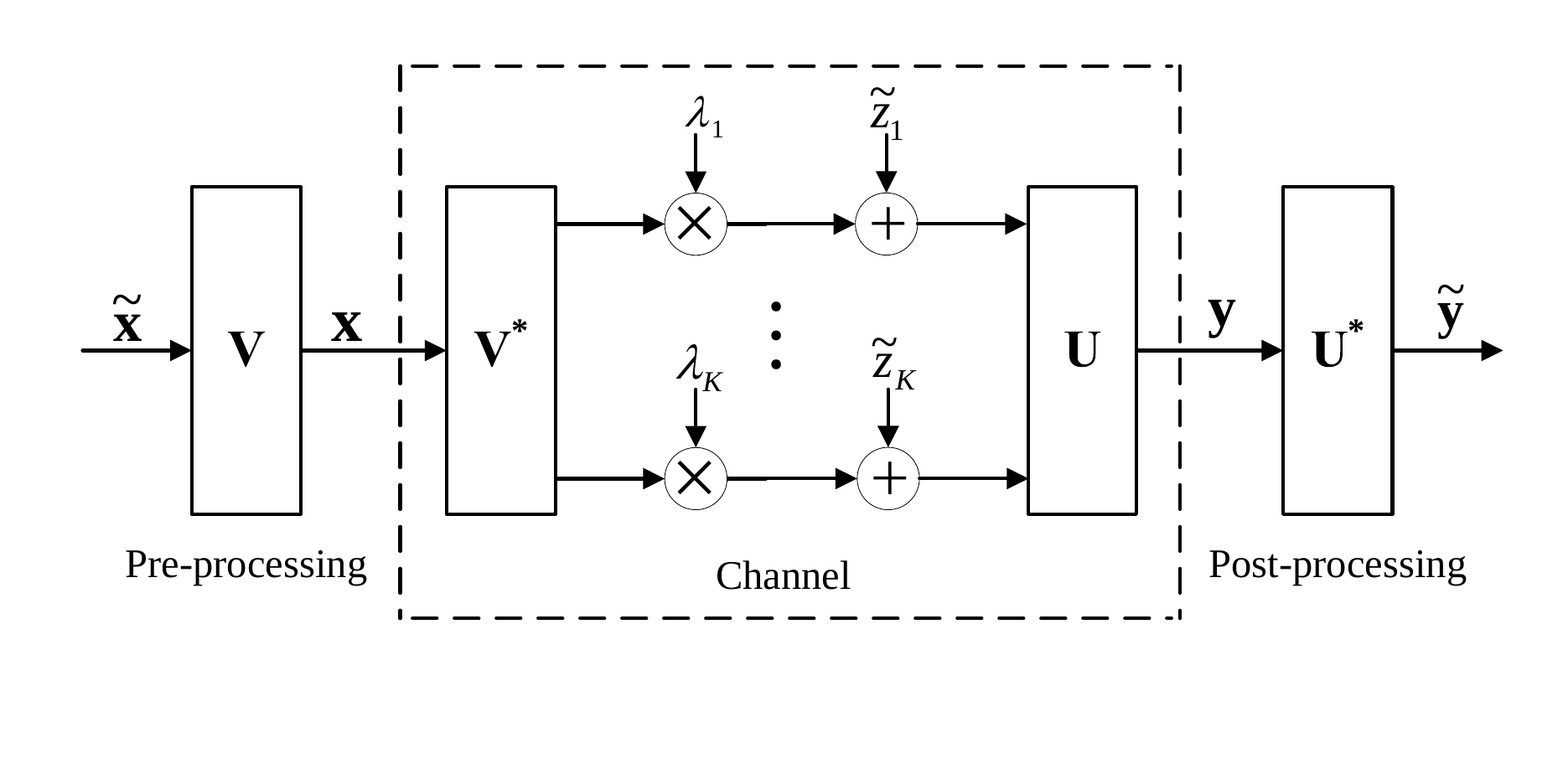}}
\small{Fig. S3. The meta-multiplexing system can be converted into
a parallel channel through singular value decomposition [1].}
\label{default}
\end{figure}

\begin{figure}[htbp]

\centerline{\includegraphics[width=9cm]{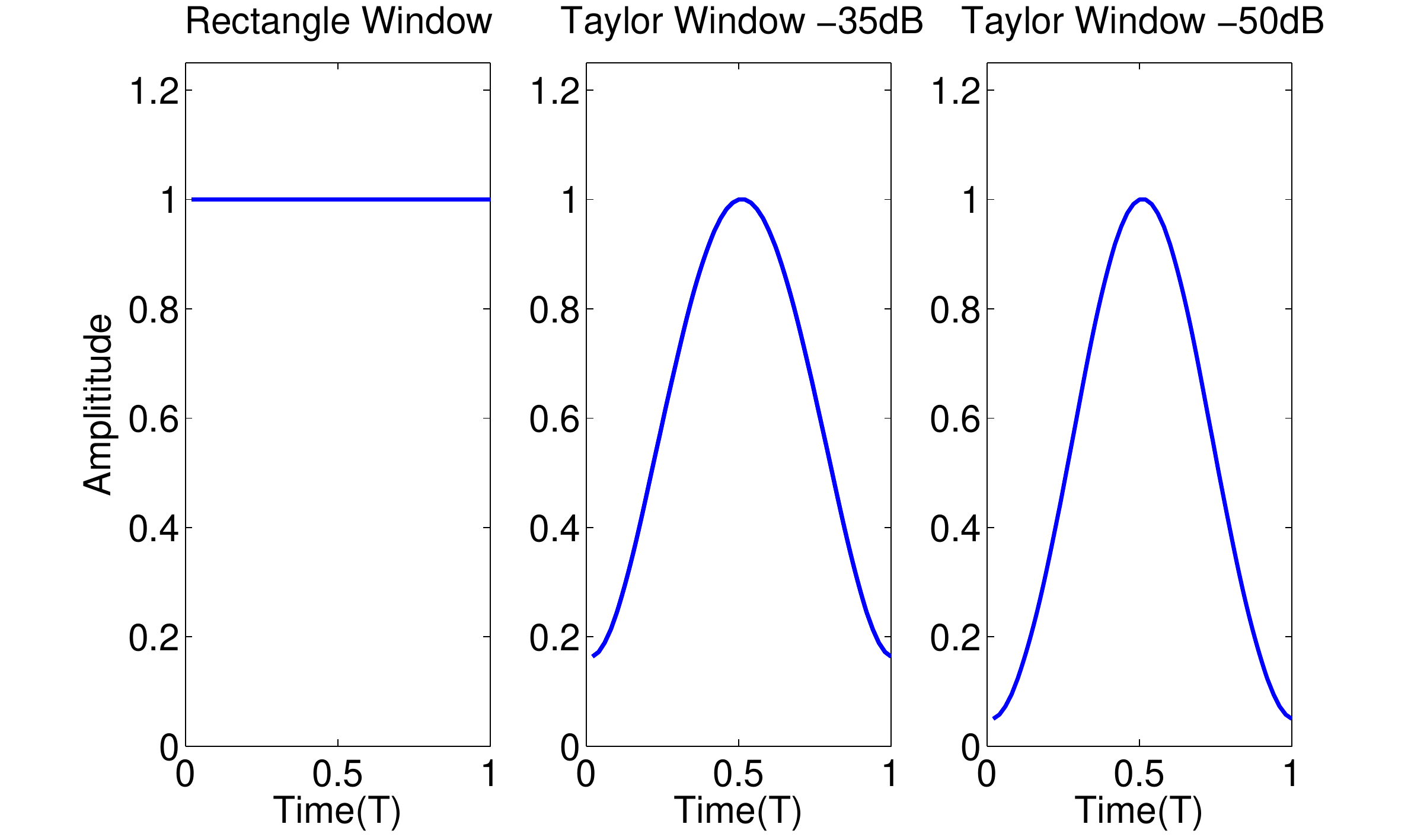}}
\centering (A)
\centerline{\includegraphics[width=9cm]{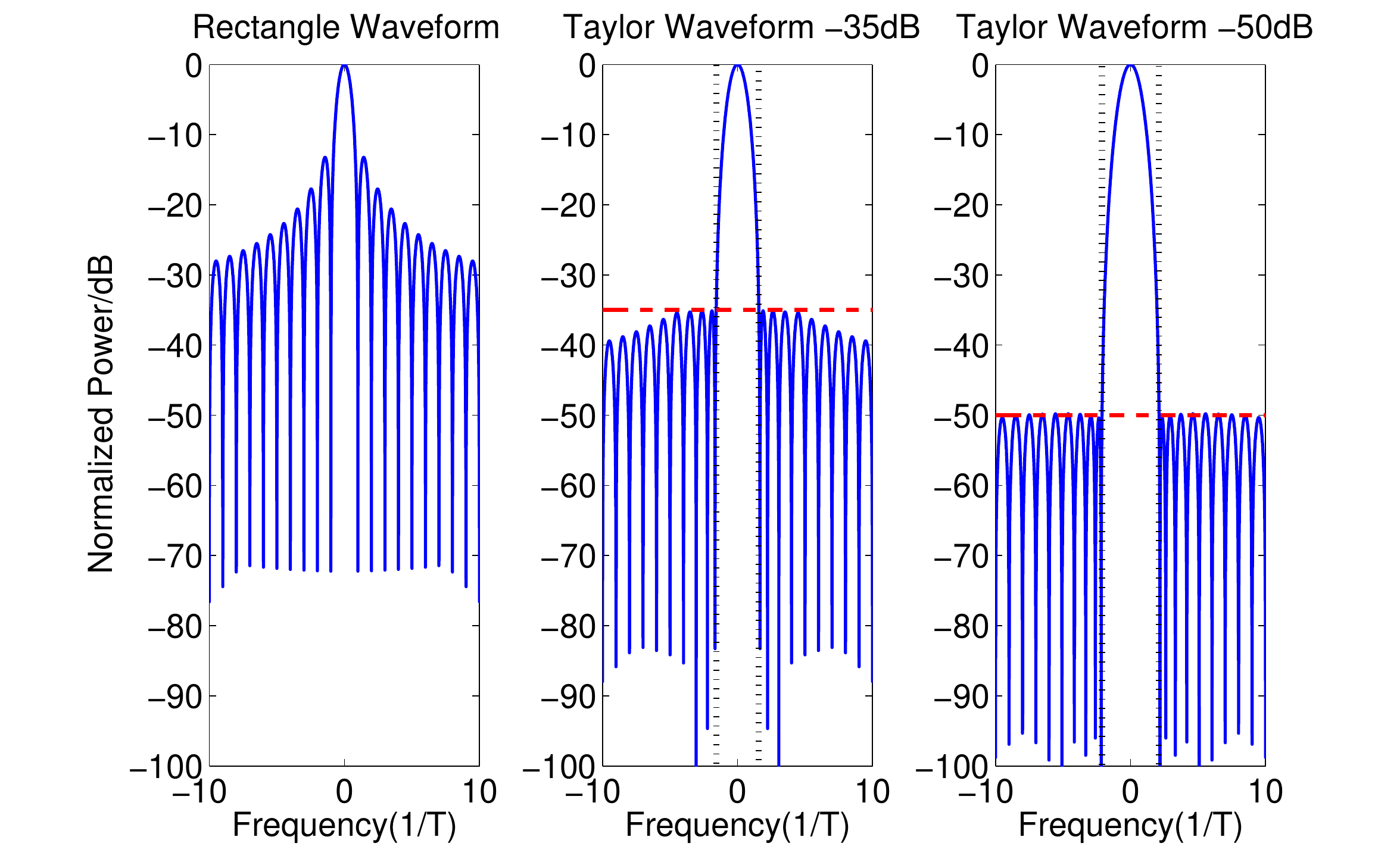}}
\centering (B)
\centerline{\includegraphics[width=9cm]{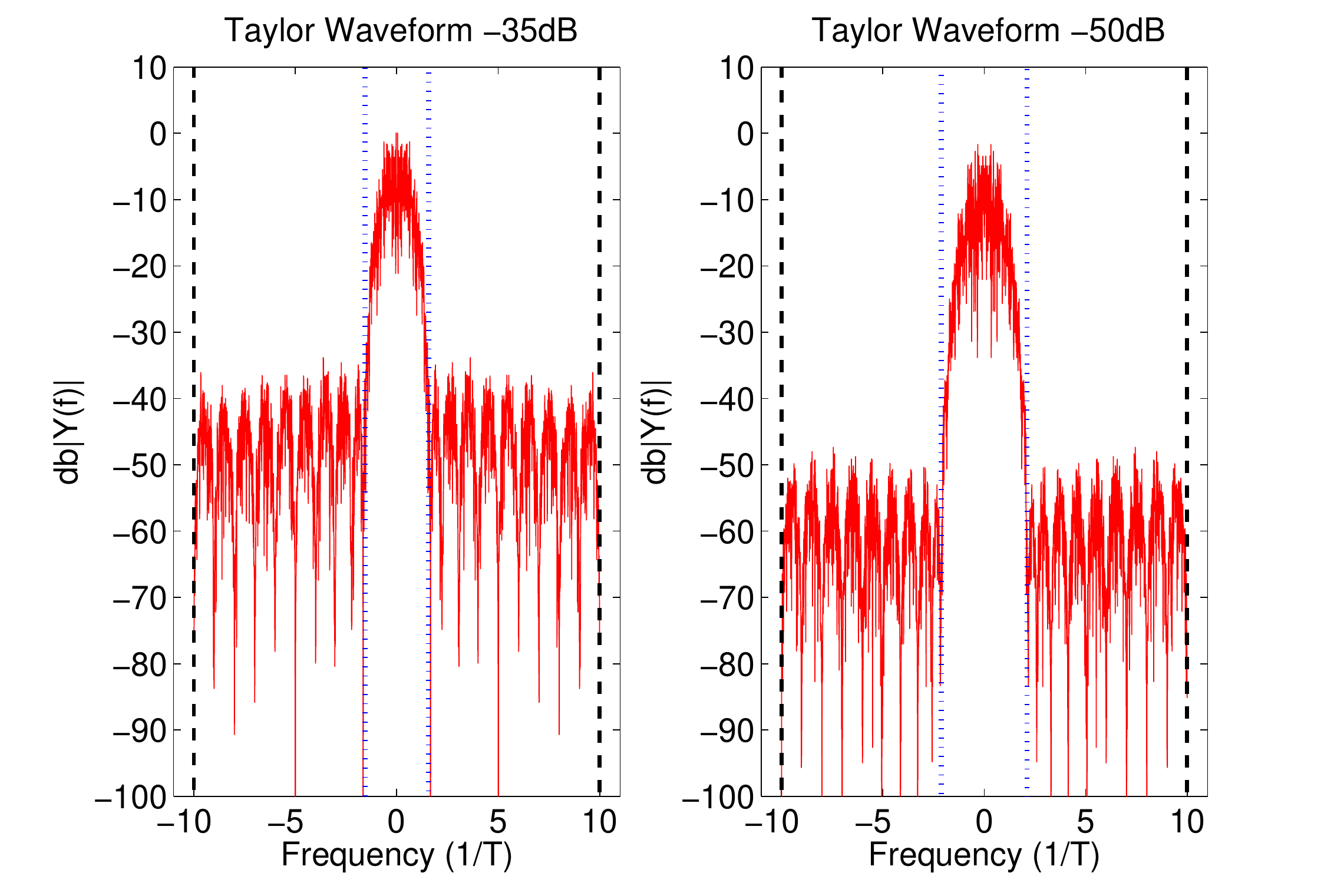}}
\centering (C)\\

\small{Fig. S4. (A) Basic waveforms in time domain. (B) Spectrum
of the basic waveforms; dotted lines show the bound PSD bandwidth
of Taylor waveforms. (C) Spectrum of the meta-multiplexing signal
with and data length; dotted lines show the bound PSD bandwidth,
while dash lines the processing bandwidth. } \label{default}
\end{figure}

\begin{figure}[htbp]
\centerline{\includegraphics[width=9cm]{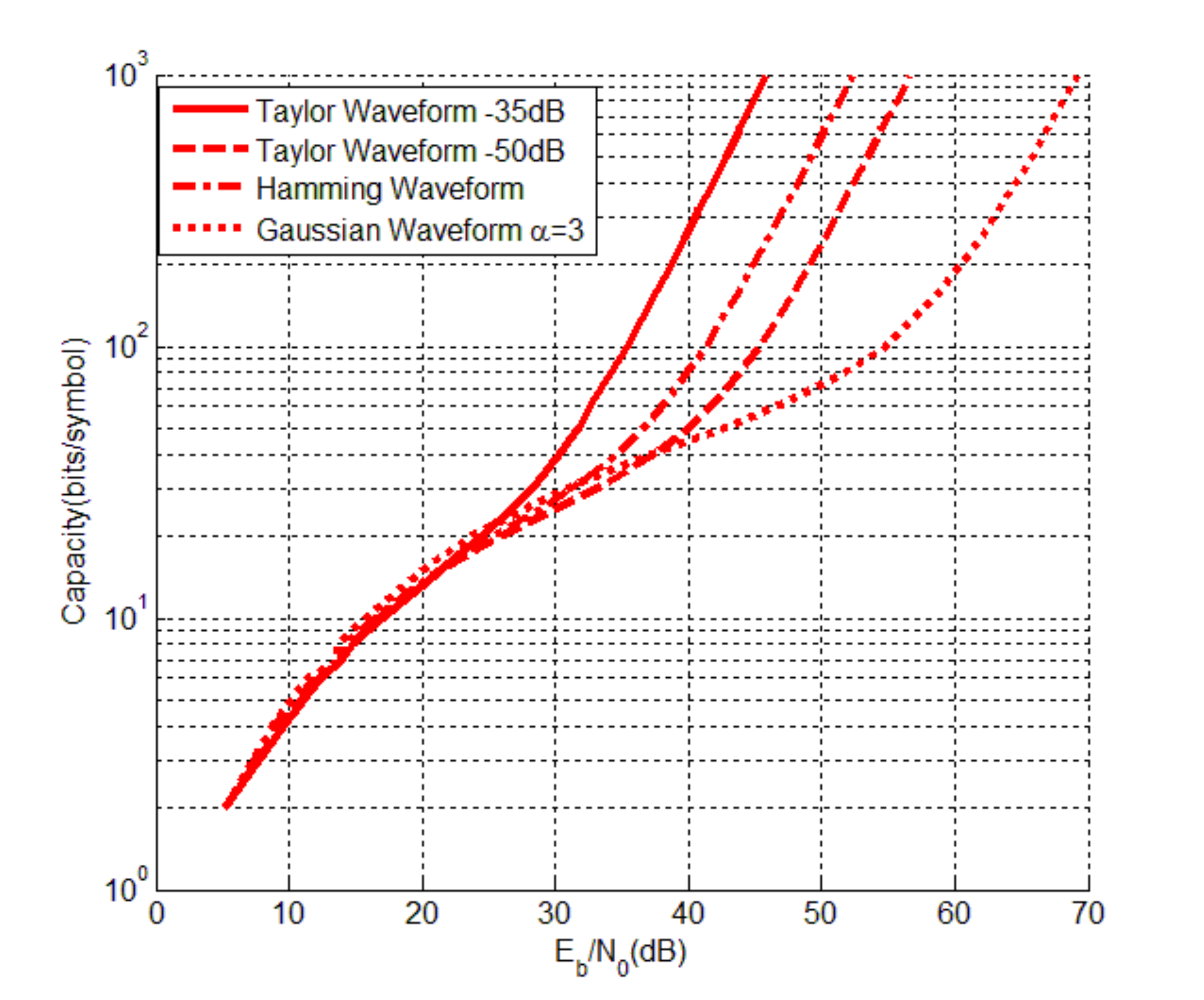}}
\centerline{\includegraphics[width=9cm]{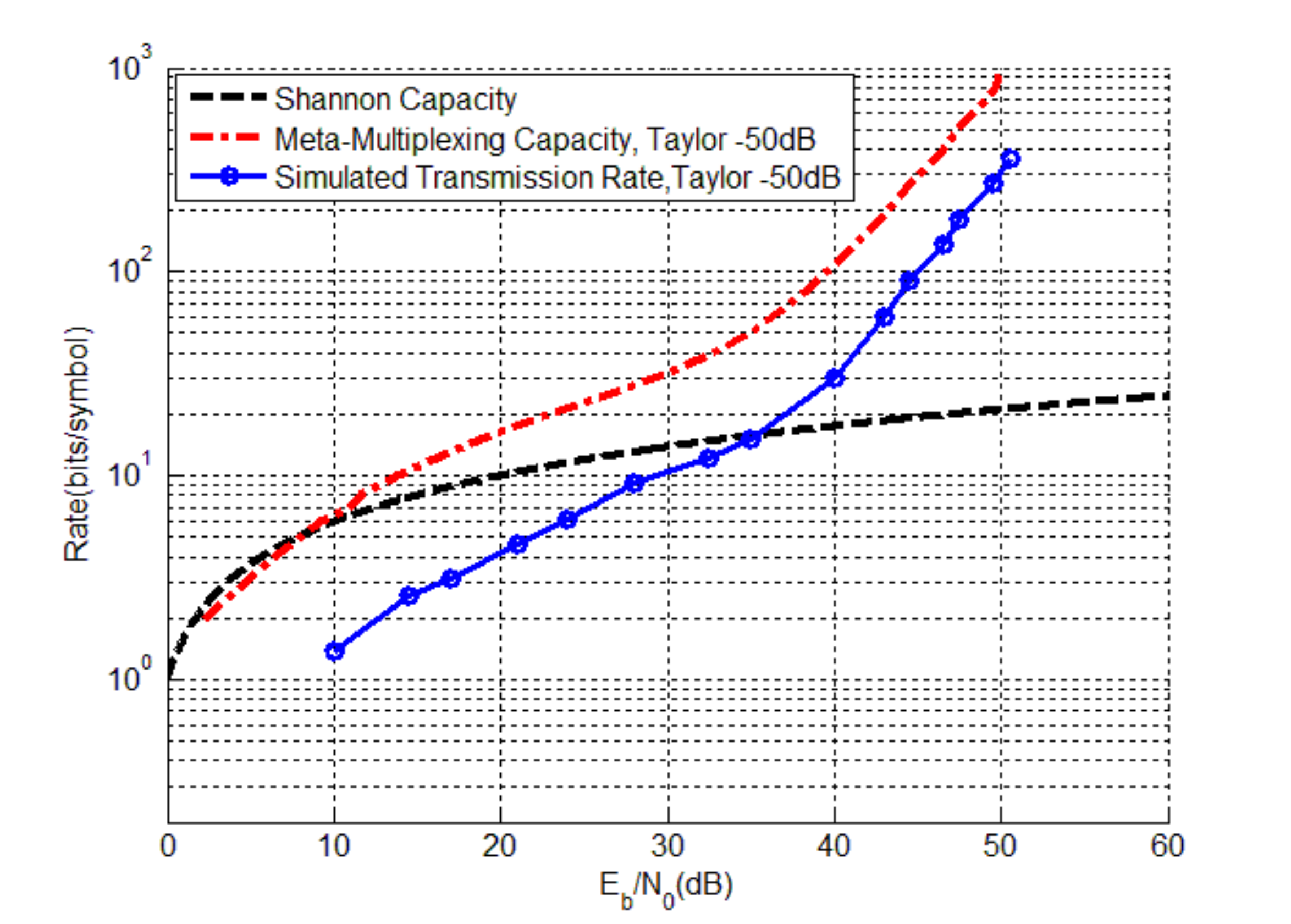}}
\small{Fig. S5. (A) The capacity of meta-multiplexing systems with
different waveforms. (B) The capacity and the realized
transmission rate of the meta-multiplexing system, which uses the
Taylor waveform with an attenuation level of -50 dB as the pulse
shaping filter. } \label{default}
\end{figure}

\begin{figure}[h]
\centerline{\includegraphics[width=9cm]{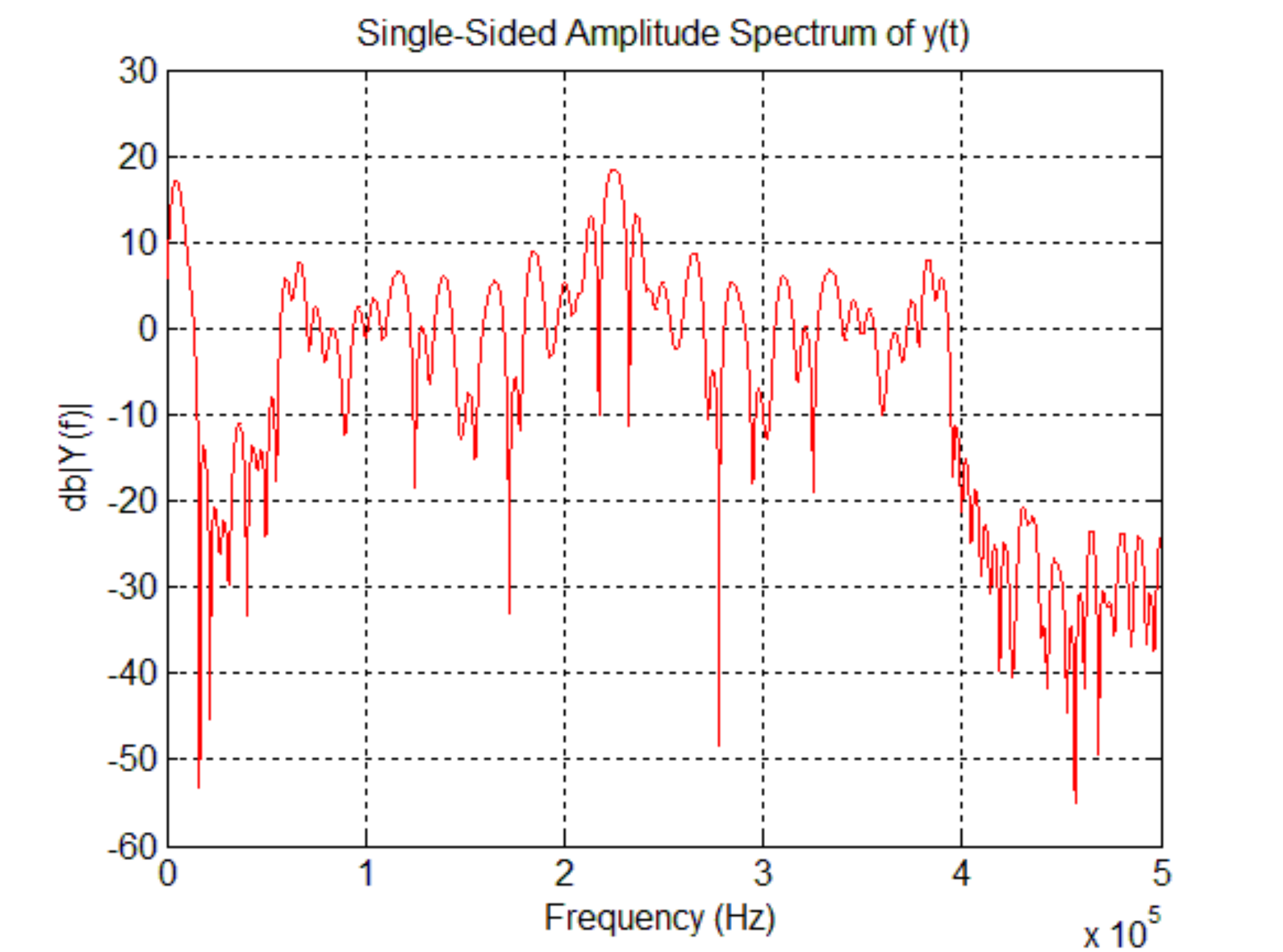}}
\small{Fig. S6. Plot of the spectrum sharing scheme. The
meta-multiplexing use an overlap factor K=100 and a Taylor
waveform with an attenuation level of -35 dB. One 256-QAM signal
occupies approximately 75\% of the processing bandwidth of the
meta-multiplexing signal.} \label{default}
\end{figure}

\begin{figure}[h]
\centerline{\includegraphics[width=12cm]{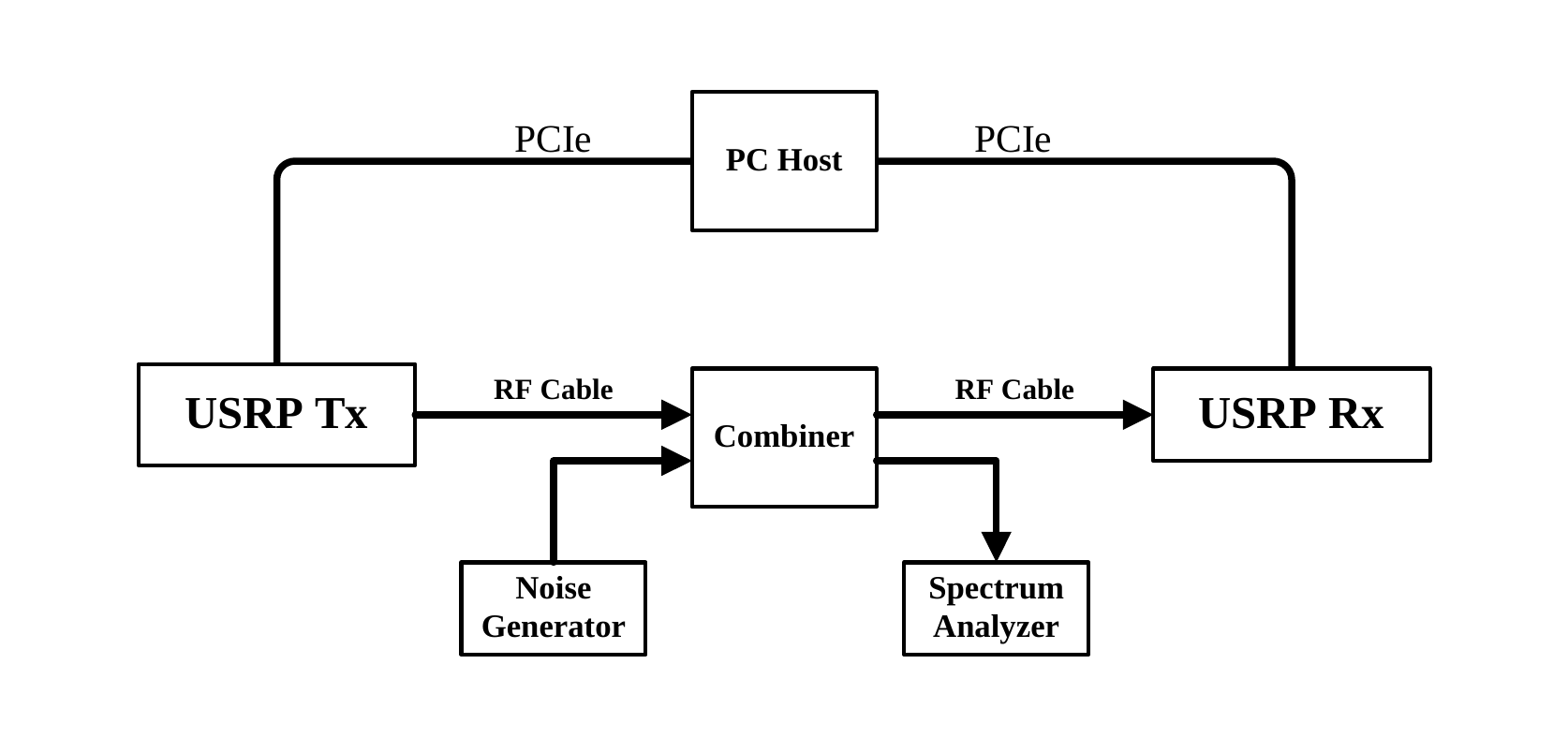}} \small{
Fig. S7. The connectivity of the USRP devices for hardware
verification.}
\end{figure}

\begin{figure}
\centerline{\includegraphics[width=10cm]{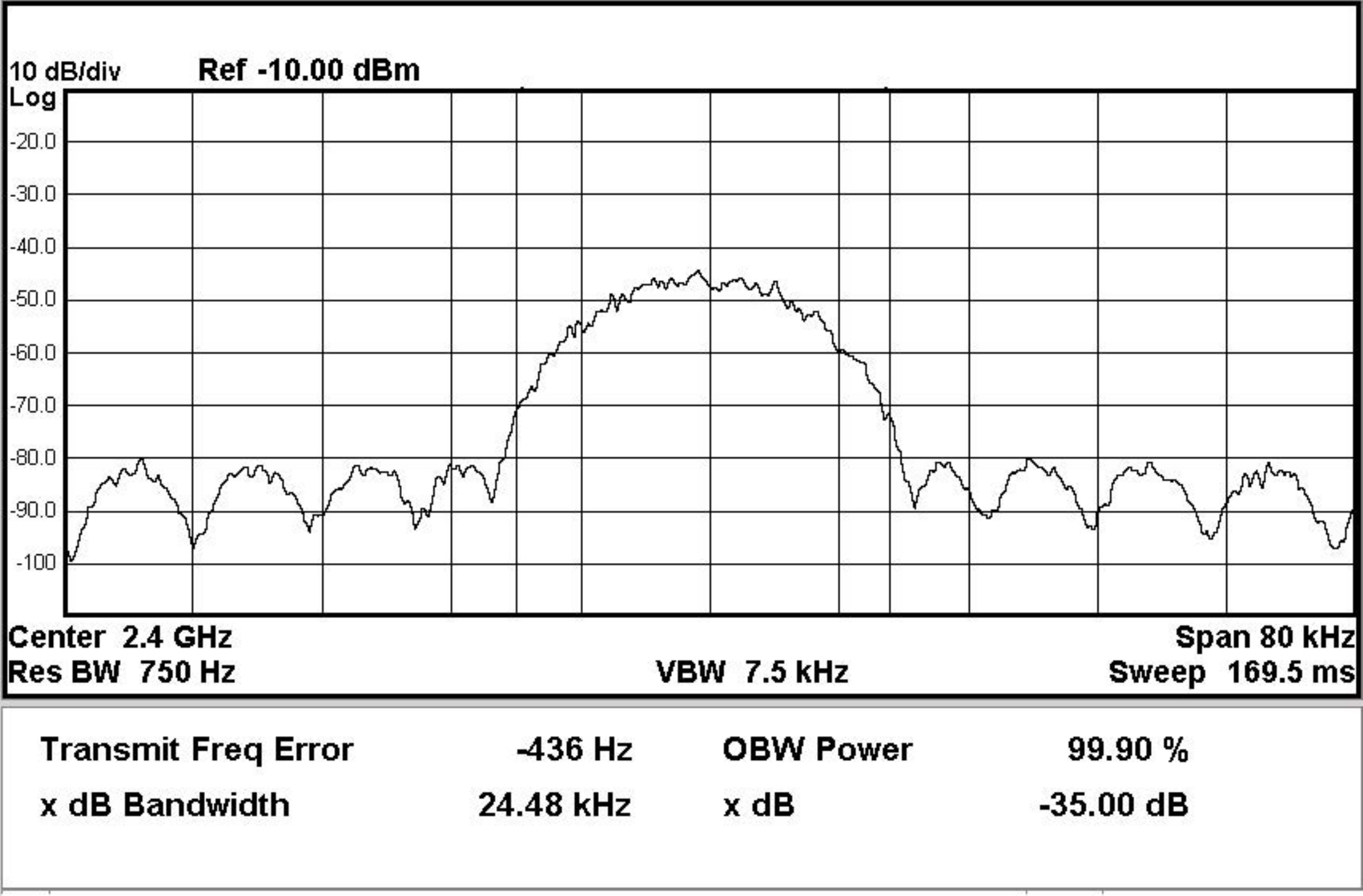}}
{\small{Fig. S8. Measure the frequency response of the
meta-multiplexing signal in the real physical channel by a
spectrum analyzer. The signal is rolled off to -35dB in a
bandwidth of 24.48 KHz.}}
\end{figure}

\end{document}